\documentclass{article}
\usepackage{arxiv}
\usepackage[utf8]{inputenc} 
\usepackage{graphicx}
\usepackage{subcaption}
\usepackage[T1]{fontenc}    
\usepackage{amsmath}
\usepackage{amsfonts}       
\usepackage{nicefrac}       
\usepackage{microtype}      
\usepackage{lipsum}
\usepackage{xcolor}
\usepackage{dcolumn}
\usepackage{bm}
\usepackage{array}
\usepackage{cite}
\DeclareMathAlphabet{\pazocal}{OMS}{zplm}{m}{n}
\begin{document}

\title{Maximally modular structure of growing hyperbolic networks}

\author{
  S\'{a}muel G. Balogh\\
  Dept. of Biological Physics, 
  E\"{o}tv\"{o}s University,
  H-1117 Budapest, Pázmány P. stny. 1/A, Hungary\\
  \texttt{*balogh@hal.elte.hu} \\
   \AND
 Bianka Kovács \\
  Dept. of Biological Physics, 
  E\"{o}tv\"{o}s University,
  H-1117 Budapest, Pázmány P. stny. 1/A, Hungary\\
   \AND
   Gergely Palla \\
  Dept. of Biological Physics, Eötvös Loránd University, H-1117 Budapest, Pázmány P. stny. 1/A, Hungary\\
MTA-ELTE Statistical and Biological Physics Research Group, H-1117 Budapest, Pázmány P. stny. 1/A, Hungary\\
Health Services Management Training Centre, Semmelweis University,  H-1125, Kútvölgyi út 2, Budapest, Hungary
}


\date{\today}

\maketitle

\begin{abstract}
Hyperbolic models are remarkably good at reproducing the scale-free, highly clustered and small-world properties of networks representing real complex systems in a very simple framework. Here we show that for the popularity-similarity optimization model from this family, the generated networks become also extremely modular in the thermodynamic limit, in spite of lacking any explicit community formation mechanism in the model definition. According to our analytical results supported by numerical simulations, when the system size is increased, the modularity approaches one surprisingly fast.

\end{abstract}

\vspace{0.3cm}
Networks representing the patterns of interactions between the fundamental units of complex systems can show immensely 
rich behaviour as demonstrated by a vast number of studies, forming the core subject of an interdisciplinary field that became widely popular in the last two decades~\cite{Dorog_book,Laci_revmod,Newman_Barabasi_Watts,Jari_Holme_Phys_Rep,Vespignani_book}. The most important features of complex networks that show a great deal of universality across systems ranging from the metabolic networks within cells to the level of the entire society are the inhomogeneous scale-free nature of the degree distribution~\cite{Faloutsos,Laci_science}, the high local transitivity characterised by a relatively large average clustering coefficient~\cite{Watts-Strogatz}, and the small-world property~\cite{Milgram_small_world,Kochen_book}. Furthermore, most real networks also display an intricate community structure~\cite{Fortunato_coms,Fortunato_Hric_coms,Cherifi_coms}, corresponding to the presence of denser modules in the network topology, in a similar fashion to families and friendship circles in the society. Capturing the most essential properties of complex networks with the help of simple mathematical models has always been one of the key goals in this field, and a notable approach in this respect is given by hyperbolic models~\cite{hyperGeomBasics,PSO,EPSO_HyperMap,GPA_PSOsoftComms,nPSO,S1,S1softComms,dPSO}, centred around the idea of placing the nodes in hyperbolic space and connecting node pairs with a probability depending on the hyperbolic distance. 

One of the first models based on this idea is the popularity-similarity optimization (PSO) model~\cite{PSO}, working in the native disk representation of the two-dimensional hyperbolic space, where $N$ number of nodes are introduced one by one at logarithmically increasing radii with uniformly random angular coordinates. A new-coming node $i$ is connected to the already existing nodes with a probability decaying as a function of the hyperbolic distance $x_{ij}$ as 
\begin{equation}
            p(x_{ij})=\frac{1}{1+e^{\frac{\zeta}{2T}(x_{ij}-R_i)}},
            \label{eq:PSO_con_prob}
\end{equation}
where $\zeta = \sqrt{-K}$ parametrises the curvature $K<0$ of the hyperbolic space (where usually $\zeta=1$ is used), 
$T\geq 0$ is a model parameter called temperature, and $R_i$ is the cutoff distance of the connection probability at the arrival of node $i$, 
adjusted in such a way that the expected number of links formed between the new node $i$ and the rest of the system is equal to $m\geq 1$ that acts as an additional model parameter related to the average degree. Due to the hyperbolic nature of the geometry, the radial coordinate of the nodes has a very strong effect on the degree, with the most inner nodes usually becoming hubs in the long run. In order to allow control over the degree distribution, an outward shift of the nodes is also introduced as
$r_{ji} = \beta r_{jj}+(1-\beta)r_{ii}$,
where $r_{ji}$ denotes the radial coordinate of node $j$ at the appearance of node $i$, and $\beta\in(0,1]$ is a further model parameter. 

Remarkably, the networks generated by the PSO model are small-world and scale-free (where the degree decay exponent $\gamma$ can be adjusted by the popularity fading parameter $\beta$ as $\gamma=1+1/\beta$), with a tunable clustering coefficient~\cite{PSO}. Moreover, recent numerical studies have shown that community finding methods can detect the presence of a surprisingly strong community structure in PSO networks for a wide range of the model parameters~\cite{commSector_commDetMethod,commSector_hypEmbBasedOnComms_2016,commSector_hypEmbBasedOnComms_2019,our_hyp_coms}.
In the present paper, we bring the research focusing on the community structure of hyperbolic networks to a new level by showing analytically that the modularity of PSO networks can approach 1 in the thermodynamic limit.

The modularity $Q$ corresponds to the most commonly used quality measure for quantifying the strength of communities~\cite{Newman_modularity_original,Fortunato_coms,Fortunato_Hric_coms}, comparing the observed fraction of links inside the modules with its expected value based on a null model, which is usually the configuration network ensemble. A basic form of $Q$ can be written as
\begin{equation}
Q=\sum\limits_{c=1}^{q}\left [ \frac{l_c}{E}-\left ( \frac{\sum\limits_{i \in c}k_{i}}{2E} \right )^2\right ],
\label{eq:exact_Q}
\end{equation}
where the summation runs over the communities, $l_{c}$ denotes the 
number of links inside module $c$, $k_i$ is the degree of the community member $i$, and $E$ stands for the total number of links in the network. Introducing $b_i$ as the number of intra-community links of node $i$, we can express $Q$ also as
\begin{equation}
Q=\sum\limits_{c=1}^{q}\left [ \frac{\sum\limits_{i\in c} b_i}{2E}-\left ( \frac{\sum\limits_{i \in c}k_{i}}{2E} \right )^2\right ].
\end{equation}

Since the modules of hyperbolic networks located by community finding algorithms in previous studies corresponded mostly to separated angular regions~\cite{commSector_commDetMethod,commSector_hypEmbBasedOnComms_2016,commSector_hypEmbBasedOnComms_2019,our_hyp_coms} 
(often named as angular "sectors"), here we also define the partitioning of the PSO network according to the angular node coordinates and divide the native disk into $q$ number of communities of equal angular width given by $2\pi/q$ (where we assume that $q\geq 2$ and $q<<N$, allowing at least a few members in each community). The boundary between the first and the second community can be placed at any angle $\alpha$, and once this is fixed, the rest of the community boundaries are found at $\alpha+i2\pi/q$, where $i$ runs up to $q-1$. Naturally, the modularity $Q$ depends on the chosen value of $q$ and, up to a certain variation, also on $\alpha$ -- see Sect.~S2 of the Supplementary Material (SM). However, by assuming that nodes and links are distributed among the communities evenly, $Q$ can be approximated as
\begin{equation}
  Q(q)  \approx  \frac{\left ( \sum\limits^N_{i=1} b_{i}\right)}{2E}-  q\frac{\left( \left(\sum\limits_{i=1}^N k_{i} \right )/q\right )^2}{(2E)^2}.
\end{equation}
Replacing $b_i$ with the expected number of internal links of node $i$ (denoted by $\bar{b}_i(q)$) and using that the sum of the node degrees in the negative term is equal to $2E$, the expected value of the modularity can be given as
\begin{equation}
   \bar{Q}(q) \approx \frac{1}{2E}\sum\limits^N_{i=1} \bar{b}_{i}(q)-\frac{1}{q}.
    \label{eq:av_Q_base}
\end{equation}

Following a similar line of derivation as in Ref.~\cite{EPSO_HyperMap} for the expected degree $\bar{k}_i(t)$ of node $i$ appearing at time $t=i$ as a function of $t$ during the network generation process, 
the expected internal degree for the same node, denoted by $\bar{b}_i(t)$, 
can also be calculated. The main idea is to focus only on the links that appear between node $i$ and other members of the community of node $i$ by replacing the connection probability with a conditional probability conditioned on that the other node falls into the same angular region as node $i$, resulting in
\begin{equation}
\bar{b}_i(t)\approx    \bar{k}_i(t)-qm^2\frac{\tan(T\pi)}{4\pi T}\frac{\left(\left(\frac{i}{t}\right)^{1-2\beta}-1\right)}{i(2\beta-1)I^2_t}
\label{eq:b_i_t}
\end{equation}
at any $T<1/2$, where $I_t=\frac{1-t^{-(1-\beta)}}{1-\beta}$ and the details of the calculation are moved to Section~S3.2.2 of the SM. Naturally, we are interested in the result for $\bar{b}_i(t)$ at the end of the network generation process where $t=N$, and the above approximation (similarly to the result for $\bar{k}_i(t)$ in Ref.~\cite{EPSO_HyperMap}) works best for $N\rightarrow\infty$. Substituting Eq.~(\ref{eq:b_i_t}) into Eq.~(\ref{eq:av_Q_base}) yields
\begin{equation}
    \bar{Q}(q) \approx 1-C_1q-\frac{1}{q},
    \label{eq:avg_mod_q_dependence}
\end{equation}
where $C_1=C_1(N,m,T,\beta)$ is independent of the number of communities $q$, and can be written as
\begin{equation}
C_1=\frac{m}{2(2\beta-1)NI^2_N}\frac{\tan(\pi T)}{4\pi T}\sum\limits^{N}_{i=1}\frac{1}{i}\left ( \left ( \frac{i}{N} \right )^{1-2\beta} -1\right ).
\label{eq:C_1_base}
\end{equation}

\begin{figure}[hbt]
    \centering
\includegraphics[width=0.75\textwidth]{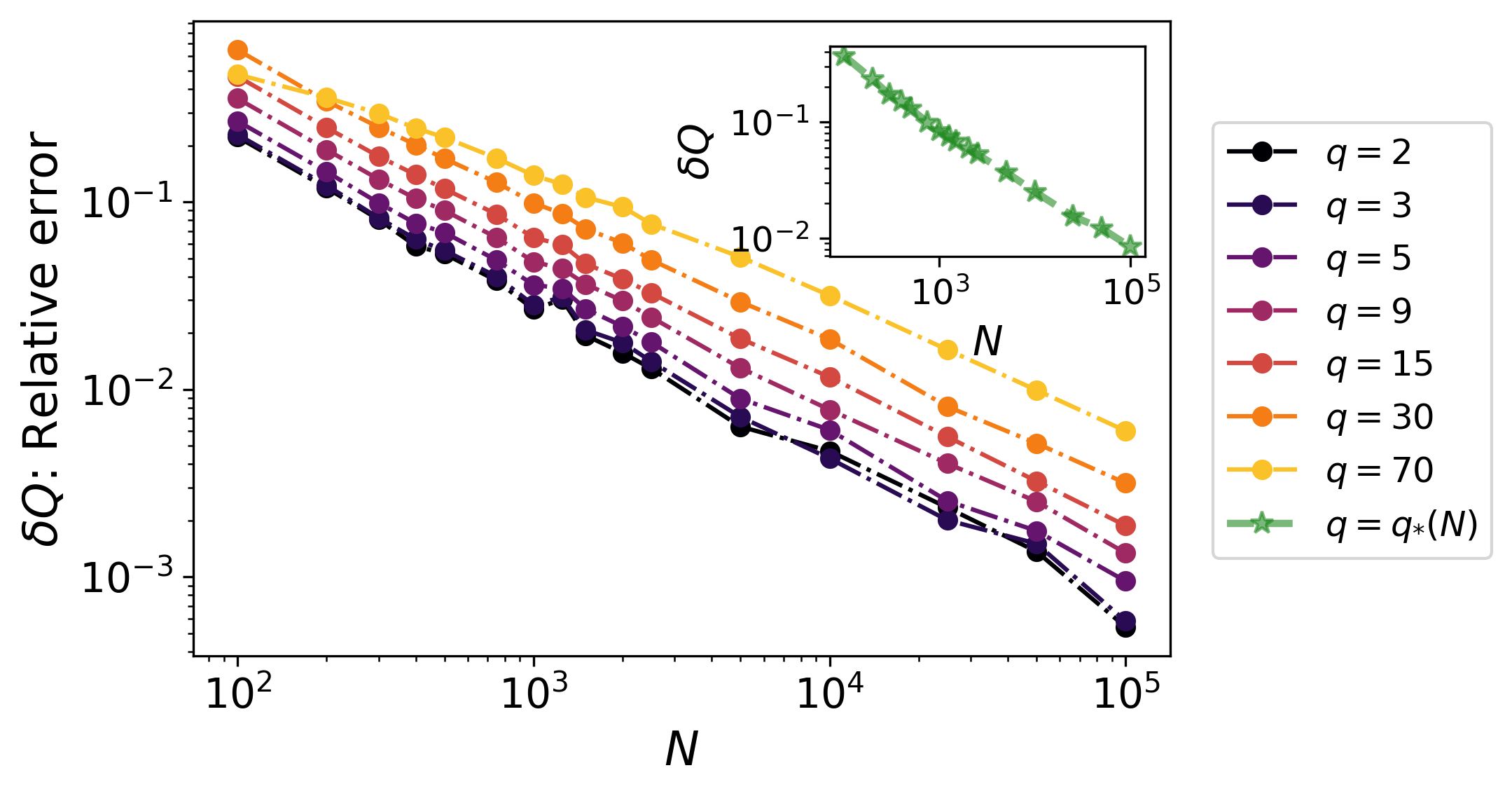}
\caption{\label{fig:errors} {\bf The relative error of the analytic modularity.} The expected modularity $\bar{Q}(q)$ from Eq.~(\ref{eq:avg_mod_q_dependence}) is compared to the exact modularity $Q(q)$ calculated according to Eq.~(\ref{eq:exact_Q}), and averaged over PSO networks generated with parameters $m=2,\,\beta=0.6,\,T=0.1$, when the networks are divided into $q$ equally sized communities according to the angular coordinates of the network nodes. 
The relative error is defined as $\delta Q=[\bar{Q}(q)-\left< Q(q)\right>]/\left< Q(q)\right>$. The number of samples considered in $\left< Q(q)\right>$ 
decreases with $N$ from $10000$ at ${N=10^2}$ to $50$ at $N=10^5$. The inset shows $\delta Q$ at the optimal ($N$-dependent) $q_*$ value, where the expected modularity is maximal.}
\end{figure}

Before discussing the consequences of the above results, let us examine how well Eq.~(\ref{eq:avg_mod_q_dependence}) approximates the modularity according to the relative error $\delta Q$ with respect to the average modularity measured in networks generated by the PSO model with uniform angular partitioning into $q$ communities. In Fig.~\ref{fig:errors}, we show $\delta Q$ as a function of the network size $N$ at different $q$ values and fixed $m,\,\beta,\,T$ parameters, displaying a clear decreasing tendency. The similar behaviour of the absolute error and the results at different $\beta$ and $T$ parameters are presented in Sect.~S3.5 of the SM (together with an analysis of the variation of the measured $Q$ across different network samples and different starting angles defining the community boundaries in the native disk). These results suggest that Eq.~(\ref{eq:avg_mod_q_dependence}) becomes 
exact in the $N\rightarrow \infty$ thermodynamic limit. 

Based on Eq.~(\ref{eq:avg_mod_q_dependence}), the determination of the $q=q_*$ value where $\bar{Q}(q)$ is maximal is straightforward by taking the derivative with respect to $q$, leading to $q_{*}=C^{-1/2}_1$. 
By substituting back into Eq.~(\ref{eq:avg_mod_q_dependence}), we obtain that the maximal value of the modularity is 
\begin{equation}
    \bar{Q}(q_{*}) = 1-2C^{1/2}_1.
    \label{eq:Q_at_opt_q}
\end{equation}
In the inset of Fig.~\ref{fig:errors}, we show the relative error of $\bar{Q}(q_*)$ based on Eq.~(\ref{eq:Q_at_opt_q}), displaying a decreasing tendency with $N$ in a fashion similar to the main plot.

Besides $q_*$, another important $q$ value is given by the resolution limit of the modularity~\cite{Fortunato_Barthelemy_resolution_limit_2007}, where the communities become too small compared to the system size, and the number of internal links in a single module drops below $\sqrt{E/2}$. 
In such a case the partitioning is so far from optimal that merging any pairs of communities that are connected by at least a single link will increase $Q$. Using Eq.~(\ref{eq:b_i_t}), the expected number of internal links in each community as a function of $q$ can also be simply calculated (the details are given in Sect.~S5 of the SM), yielding 
at the resolution limit 
\begin{equation}
q_{\rm res}  = \frac{\sqrt{2E}}{1+C_1 \sqrt{2E}}= \frac{q_*^2}{1+\frac{q^2_{*}}{\sqrt{2E}}}.
\end{equation}

\begin{figure}[hbt]
\centering
\includegraphics[width=0.6\textwidth]{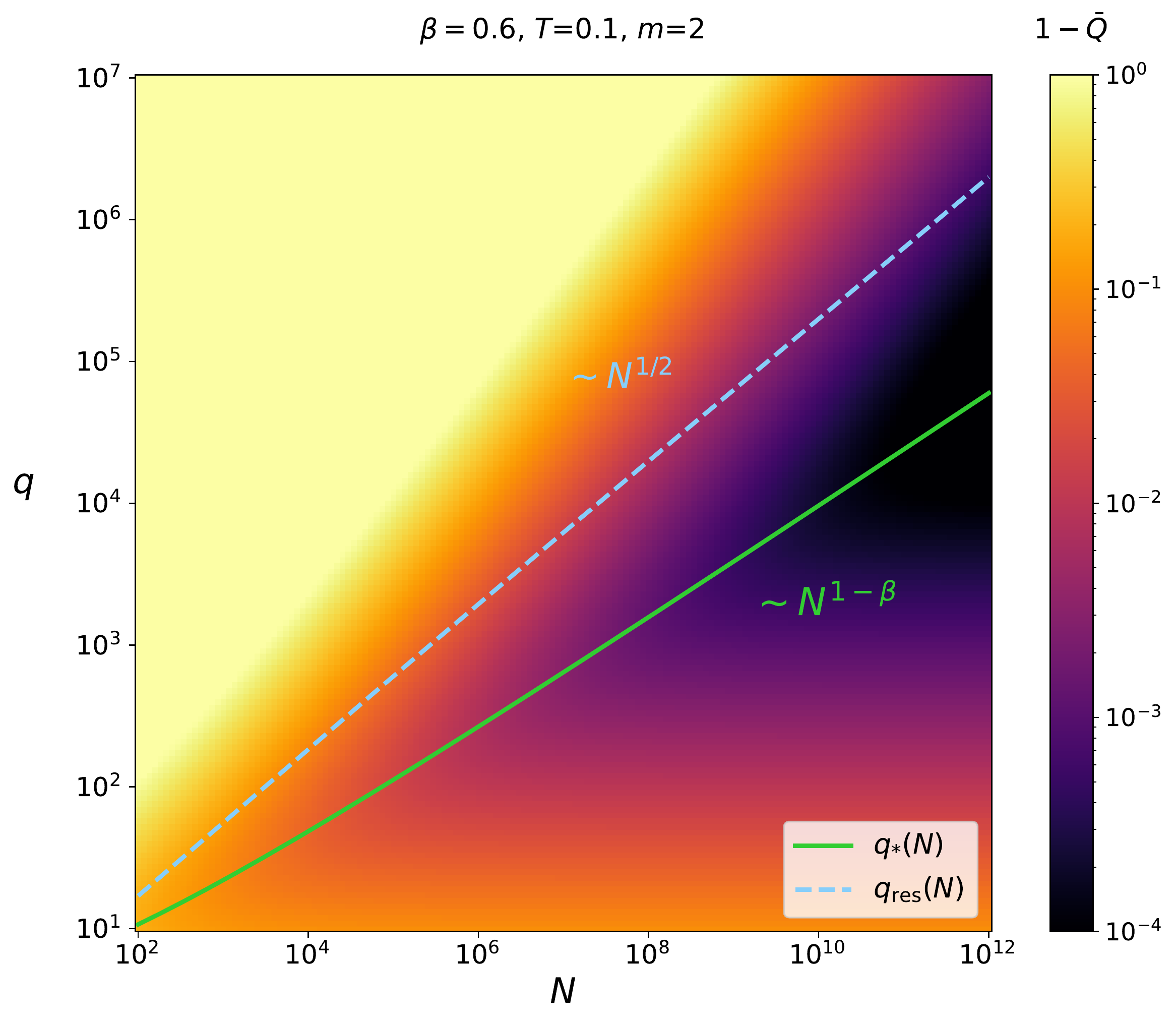}
\caption{\label{fig:heatmap} {\bf The expected modularity as a function of $N$ and $q$.} For better visibility, we plot $1-\bar{Q}$ with the help of the colormap, 
showing that $\bar{Q}$ can get very close to 1 already in the examined network size range. The continuous line corresponds to $q_*(N)$, whereas the dashed line shows $q_{\rm res}(N)$.}
\end{figure}

Let us now turn to the behaviour of the modularity itself by plotting $1-\bar{Q}(q)$ in Fig.~\ref{fig:heatmap} as a function of both $N$ and $q$, where $\bar{Q}(q)$ is obtained from Eq.~(\ref{eq:avg_mod_q_dependence}). 
The heat-map clearly indicates that the expected modularity approaches $1$ if $N$ is increased and $q$ is in the vicinity of $q_*(N)$. According to the calculations, the modularity of PSO networks can surpass $0.9$ already at $N=10^4$, and even $0.99$ at $N=10^6$, which is smaller only by 1\% compared to the theoretically possible maximum value of $1$.

\begin{figure}[hbt]
    \centering
    \includegraphics[width=0.85\textwidth]{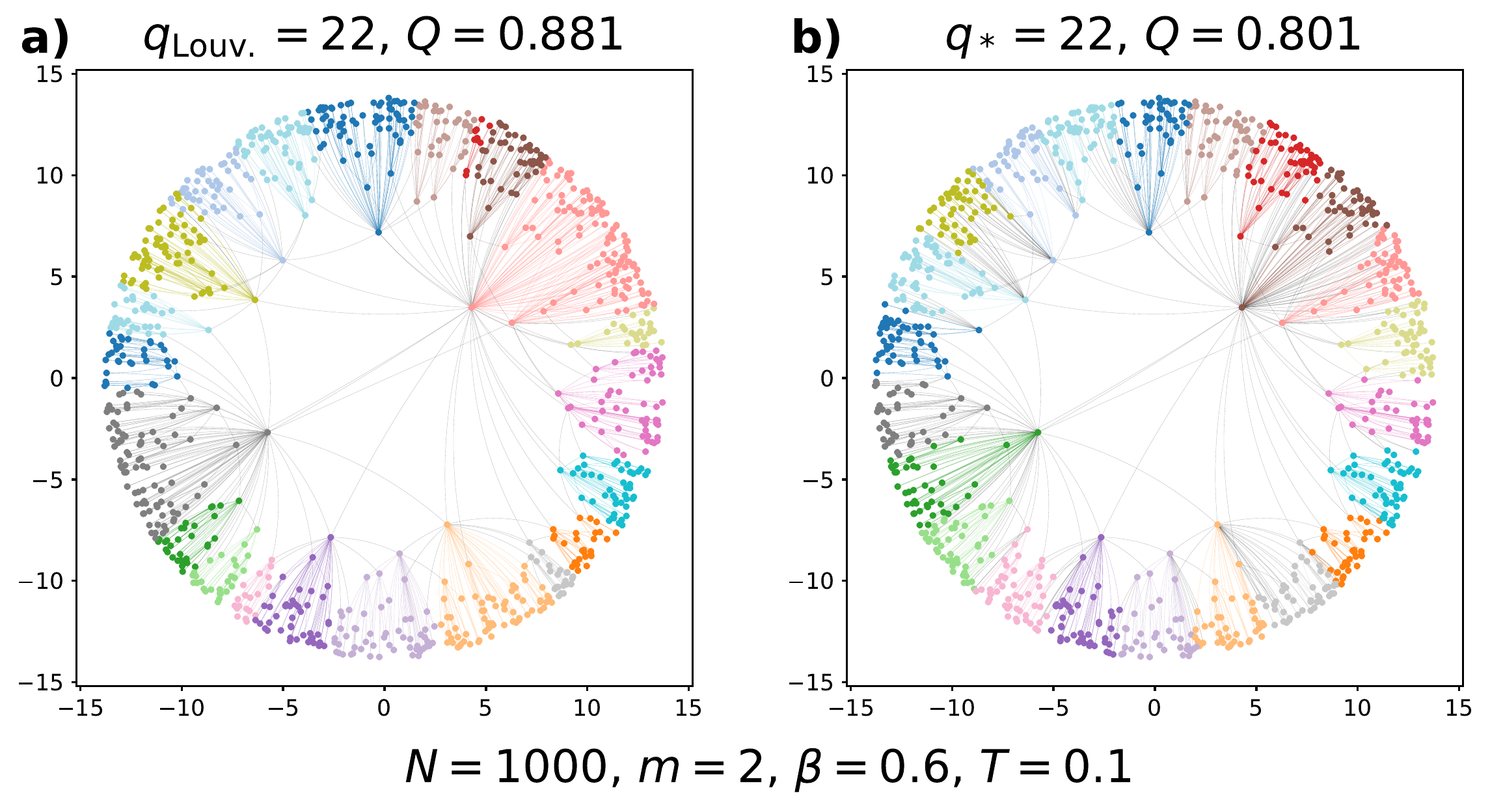}
    \caption{ {\bf Communities in a PSO network.} 
    a) The modules found by the Louvain algorithm, indicated by colour. b) The communities according to the partitioning studied in this paper, consisting of equally sized angular regions. The number of modules in this case was set to the optimal $q_*$ calculated as $q_{*}=C^{-1/2}_1$, using Eq.~(\ref{eq:C_1_base}).}
    \label{fig:layouts}
\end{figure}

Next, as an illustration, 
we show in Fig.~\ref{fig:layouts}a a PSO network of ${N=1000}$ number of nodes with the communities located by the Louvain algorithm~\cite{Louvain} (corresponding to a very popular community finding method built on modularity maximisation), compared to the partitioning of the same network in Fig.~\ref{fig:layouts}b according to the setup studied here, where the communities are defined by $q_*$ number of angular sectors of equal size. Naturally, the modules found by Louvain have varying sizes, and in general, the total number of communities can also be different in the two cases. Meanwhile, the overall look of the two partitionings is quite similar, and the modularity of the setup with $q_*$ equally sized communities is close to the modularity of the partitioning found by Louvain, showing that the proposed division of PSO networks into communities is not far from optimal already at such small network sizes. We note however that the true optimum of the modularity is expected to be higher compared to the $Q$ measured for our uniform partitioning at finite network sizes, since angular sectors of varying sizes may adapt better to fluctuations in the network structure.  

An interesting remaining question is how does the modularity behave in the asymptotic $N\rightarrow\infty$ limit? According to Eq.~(\ref{eq:avg_mod_q_dependence}), for a fixed $q$ value 
\begin{equation}
    \lim\limits_{N\to\infty}\bar{Q}(q) = 1-\frac{1}{q},
    \label{eq:avg_mod_at_fix_q}
\end{equation}
where we used that $C_1$ approaches 0 when $N\rightarrow\infty$. Since the number of communities $q$ is not bounded for infinitely large PSO networks, the above equation already shows that the modularity of PSO networks can get arbitrarily close to 1 in the thermodynamic limit.

Naturally, instead of working with a fixed $q$ when $N\to\infty$, it is a better idea to consider the communities obtained at the optimal $q_*(N)$ when seeking the maximal modularity. Based on Eqs.~(\ref{eq:Q_at_opt_q}) and (\ref{eq:C_1_base}), the rate at which $\bar{Q}(q_*)$ approaches 1 is $\beta$-dependent. For simplicity, we move the details of the calculations into Sect.~S4 of the SM, and summarise the scaling of a few quantities of interest (including $1-\bar{Q}(q_*)$) in Table~\ref{tab:scaling-table}. According to the results, the modularity at $q_*$ 
converges to 1 as fast as $\left(\frac{\ln N}{N}\right)^{\frac{1}{2}}$ when $\beta<\frac{1}{2}$, and still as fast as $N^{\beta -1}$ when $\beta>\frac{1}{2}$. Interestingly, the modularity at $q_{\rm res}$ corresponding to the resolution limit also approaches 1 if $\beta<\frac{3}{4}$; however, always at a slower rate compared to $Q(q_*)$. 

\begin{table}[hbt]
\centering
\resizebox{0.75\textwidth}{!}{
\setlength{\extrarowheight}{8pt}
\begin{tabular}{|c|cccc|c|}
\hline
& \multicolumn{1}{c|}{$\beta \in \left(0,\frac{1}{2}\right)$}  & \multicolumn{1}{c|}{$\beta=\frac{1}{2}$} & \multicolumn{1}{c|}{$\beta \in \left(\frac{1}{2},\frac{3}{4}\right]$} & $\beta \in \left(\frac{3}{4},1\right)$ & $\beta=1$   \\ [1.5ex] \hline\hline

$C_{1}(N)$          & \multicolumn{1}{c|}{$\frac{\ln N}{N}$ } & \multicolumn{1}{c|}{$\frac{(\ln N)^2}{N}$} & \multicolumn{2}{c|}{$N^{2\beta-2}$}                                             & $(\ln N)^{-2}$   \\ [2ex] \hline\hline

$q_{*}(N)$          & \multicolumn{1}{c|}{$\left(\frac{N}{\ln N}\right)^{\frac{1}{2}}$ } & \multicolumn{1}{c|}{$\frac{N^{\frac{1}{2}}}{\ln N}$} & \multicolumn{2}{c|}{$N^{1-\beta}$}                                             & $\ln N$   \\ [2ex] \hline
$q_{\text{res}}(N)$ & \multicolumn{3}{c|}{$N^{\frac{1}{2}}$}                                                                                                                                                                          & $N^{2-2\beta}$                     & $(\ln N)^2$        \\ [1.5ex] \hline \hline

$1-\bar{Q}\left(q_{*}\right)$          & \multicolumn{1}{c|}{$\left(\frac{\ln N}{N}\right)^{\frac{1}{2}}$ } & \multicolumn{1}{c|}{$\frac{\ln N}{N^{\frac{1}{2}}}$} & \multicolumn{2}{c|}{$N^{\beta-1}$}                                             & $(\ln N)^{-1}$   \\ [2ex] \hline 
$1-\bar{Q}\left(q_{\text{res}}\right)$ & \multicolumn{1}{c|}{$\frac{\ln N}{N^\frac{1}{2}}$} & \multicolumn{1}{c|}{$\frac{\ln^2 N}{N^\frac{1}{2}}$} & \multicolumn{1}{c|}{$N^{2\beta-\frac{3}{2}}$}                                                          &         \multicolumn{2}{c|}{$o(1)$}   \\ [1.5ex] \hline 
\end{tabular}
}
\vspace*{5mm}
\caption{ {\bf Asymptotic scaling of the modularity.}  
We list the scaling of $C_1(N)$, $q_{*}(N)$, $q_{\text{res}}(N)$, $1-\bar{Q}(q_{*})$ and $1-\bar{Q}(q_{\text{res}})$ with the system size $N$ in the thermodynamic limit $N\to\infty$ for different values of the popularity fading parameter $\beta$ at any $T<0.5$ and $1\leq m$.}
\label{tab:scaling-table}
\end{table}

In conclusion, we have shown that despite lacking any intentional community formation mechanism in the network generation process, the modularity of PSO networks can converge to 1 in the asymptotic limit, and we have also provided the dependence of the convergence rate on the model parameters. Closely related results were obtained very recently for the random hyperbolic graph model~\cite{modularity_of_RHG_2021}. The authors therein prove that the static networks generated by this model exhibit modularity of $1$ with probability $1$ in the thermodynamic limit at temperature $T=0$ for degree decay exponents $\gamma>2$ and any average degree. Taken together, the possibly maximal modularity revealed by these results is a remarkable feature of hyperbolic network models, given that their overall structure is very similar to that of real systems. Although graph constructions where $Q\rightarrow 1$ in the asymptotic limit have been proposed earlier~\cite{Brandes_modularity_2008,Fortunato_resolution_limit_2007}, hyperbolic networks can achieve this while retaining a scale-free, highly clustered and small-world structure, reproducing the most important universal features of real complex systems.

Acknowledgement: The research was partially supported by the Hungarian National Research, Development and Innovation Office (grant no. K 128780, NVKP\_16-1-2016-0004),
by the European Union’s Horizon 2020 research and innovation programme under grant agreement no. 101021607,  and the Thematic Excellence Programme (Tématerületi Kiválósági Program, 2020-4.1.1.-TKP2020) of the Ministry for Innovation and Technology in Hungary, within the framework of the Digital Biomarker thematic programme of the Semmelweis University.

G.P. developed the concept of the study, S.G.B. derived the equations, S.G.B. carried out the numerical analysis, S.G.B. and B.K. prepared the figures, S.G.B., G.P. and B.K. contributed to the interpretation of the results, G.P., S.G.B. and B.K. wrote the paper. All authors reviewed the manuscript.

\bibliographystyle{unsrt}
\bibliography{MaxModPSOrefs}

\newpage
\pagebreak

\begin{center}
  \textbf{\LARGE Maximally modular structure of growing hyperbolic networks\\
  \vphantom{a}
  \\
  \Large Supplementary Material
}\\[1.5cm]
  \end{center}

\setcounter{equation}{0}
\setcounter{figure}{0}
\setcounter{table}{0}
\setcounter{page}{1}
\renewcommand{\thefigure}{S\arabic{figure}}
\renewcommand{\thetable}{S\arabic{table}}
\renewcommand{\theequation}{S\arabic{equation}}
\renewcommand{\thesection}{S\arabic{section}}

\section{Detailed description of the PSO model}

In the popularity-similarity optimisation (PSO) model~\cite{PSO} initially the network is empty, and the network nodes are placed one by one on the hyperbolic plane with increasing radial coordinates and uniformly random angular coordinates. The new node always connects to the previously appeared nodes with a linking probability that decreases as a function of the hyperbolic distance. The model works in the native representation of the hyperbolic plane of curvature $K<0$, where the hyperbolic plane is represented in the Euclidean plane by a disk of infinite radius
. The hyperbolic distance $x$ between two points located at polar coordinates $(r,\theta)$ and $(r',\theta')$ can be expressed as
\begin{equation}
    x=\frac{1}{\zeta}\cdot\mathrm{acosh}(\mathrm{cosh}(\zeta r)\,\mathrm{cosh}(\zeta r')-\mathrm{sinh}(\zeta r)\,\mathrm{sinh}(\zeta r')\,\mathrm{cos}(\Delta\theta))
    \label{eq:hypDist}
\end{equation}
from the hyperbolic law of cosines, where $\Delta\theta=\pi-|\pi-|\theta-\theta'||$ is the angular distance between the examined points and $\zeta=\sqrt{-K}$. At $\Delta\theta=\pi$, $x=r+r'$, while for $\Delta\theta=0$, $x=|r-r'|$, meaning that in this representation, the hyperbolic distance of a point from the disk centre is equal to its radial coordinate $r$, i.e. its Euclidean distance from the disk centre. 

The properties of a PSO network that can be tuned are the total number of nodes $N$, the expected average degree $\bar{k}$ (via the model parameter $m$ that corresponds to $\bar{k}/2$), the exponent $\gamma\geq 2$ of the tail of the degree distribution that decays as $\pazocal{P}(k)\sim k^{-\gamma}$ (via the popularity fading parameter $\beta\in(0,1]$ that controls the speed of the outward drift of the nodes during the network growth), and the average clustering coefficient $\bar{c}$ (via the temperature $T\in[0,1)$ that regulates how sharp the cutoff in the connection probability function is). Interpreting $m$ as the expected number of new connections per step -- as in the variant of the PSO model called $\mathrm{Model}_{2'}$ in the Supplementary Information of Ref.~\cite{PSO} --, the network growth can be realised using the following rules:
\begin{enumerate}
    \item In the $i$th step, node $i$ appears with the radial coordinate $r_{ii} = \frac{2}{\zeta}\ln i$ and an angular coordinate $\theta_i$ sampled from the interval $[0,2\pi)$ uniformly at random. (Note that due to this choice of the radial coordinate formula, changing the value of the curvature $K=-\zeta^2$ of the hyperbolic plane corresponds to a simple rescaling of all the hyperbolic distances. The usual custom is to simply set the value of $\zeta$ to $1$.)
    \item The radial coordinates of all the previous nodes $j<i$ are increased toward $r_{ii}$ as $r_{ji} = \beta r_{jj}+(1-\beta)r_{ii}$. This outward shift of the node positions is usually referred to as 'popularity fading', as it reduces the differences in the nodes' radial attractivity.
    \item The new node $i$ gets attached to the already existing nodes as follows: 
    \begin{itemize}
        \item[a)] If $T=0$, then node $i$ becomes connected to all nodes $j<i$ at a hyperbolic distance $x_{ij}$ not larger than
        \begin{equation}
            R_i = \left\lbrace \begin{array}{ll} 
            r_{ii}-\frac{2}{\zeta}\ln\left(\frac{2}{\pi}\cdot
             \frac{1-\mathrm{e}^{-\frac{\zeta}{2}(1-\beta)r_{ii}}}{m(1-\beta)}\right) & \mathrm{if}\;\; \beta < 1, \\
             r_{ii} - \frac{2}{\zeta}\ln\left(\frac{\zeta r_{ii}}{\pi\cdot m}\right) & \mathrm{if} \;\; \beta=1. \end{array} \right.
             \label{eq:cutoff_T0}
        \end{equation}
        \item[b)] If $T>0$, then node $i$ becomes connected to nodes $j<i$ with a probability depending on the hyperbolic distance $x_{ij}$ as
        \begin{equation}
            p(x_{ij})=\frac{1}{1+\mathrm{e}^{\frac{\zeta}{2T}(x_{ij}-R_i)}}, \label{eq:PSO_link_prob}
        \end{equation}
        where the cutoff distance $R_i$ can be written as
        \begin{equation}
            R_i = \left\lbrace \begin{array}{ll} 
            r_{ii}-\frac{2}{\zeta}\ln\left(\frac{2T}{\sin(T\pi)}\cdot
             \frac{1-\mathrm{e}^{-\frac{\zeta}{2}(1-\beta)r_{ii}}}{m(1-\beta)}\right) & \mathrm{if}\;\; \beta < 1, \\
             r_{ii} - \frac{2}{\zeta}\ln\left(\frac{T}{\sin(T\pi)}\cdot \frac{\zeta r_{ii}}{m}\right) & \mathrm{if} \;\; \beta=1. \end{array} \right.
             \label{eq:cutoff}
        \end{equation}
    \end{itemize}
\end{enumerate}

\section{Variation of the modularity in PSO networks divided into communities defined by equal angular regions}
\label{sect:homogen_circular_sectors}

The partitioning we consider in this work allows freedom in the choice of the boundary (placed at some angle $\alpha$) between the first and the second community, as mentioned in the main text. Naturally, when comparing the modularity measured for the communities obtained at different
$\alpha$ values, we can expect some variation in the result due to the inherent randomness of the network model under study. Similarly, variation in the measured $Q$ is expected also if $\alpha$ is kept fixed, but the PSO network is re-generated with the same model parameters. 

\begin{figure}[!ht]
\begin{center}
\includegraphics[width=0.9\textwidth]{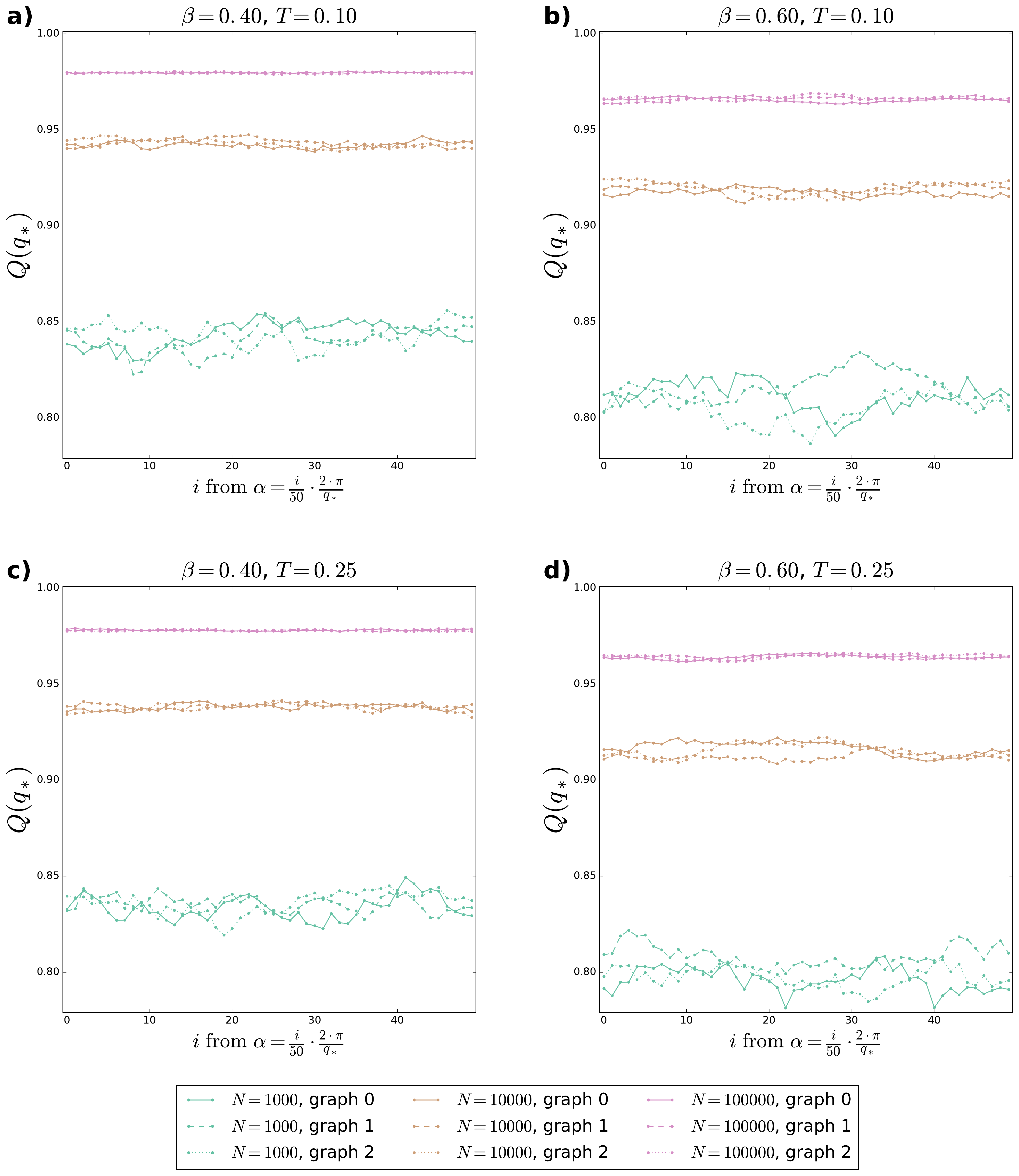}
\caption{\label{fig:Qvariation} {\bf The variation of the modularity $Q$ measured at $q_*$ number of equally sized angular sectors across different PSO network samples and different $\alpha$ starting angles defining the community boundaries in the native disk.} Each panel refers to a $\beta-T$ parameter setting given in the panel title. The curves of different colours correspond to networks of different sizes ($N=10^3,\,10^4$ or $10^5$), while the different line styles belong to different networks that were obtained from the PSO model with the given model parameters. All the studied networks were generated using $\zeta=1$ and $m=2$. We always tested 50 values of the starting angle $\alpha$ that were sampled equidistantly from the interval $[0,2\pi/q_*)$, where $2\pi/q_*$ is equal to the angular width of each one of the examined angular sectors at the optimal number of sectors.}
\end{center}
\end{figure}

In this section, we examine these fluctuations by measuring the modularity (as defined in Eq.~(2) in the main text) across different network samples and different starting angles. According to the results shown in Fig.~\ref{fig:Qvariation}, the increase in the number of nodes $N$ reduces the differences between the $Q$ values measured for different starting angles and in different networks generated by the PSO model using the same model parameters. 
This suggests that the communities we defined by simply dividing the PSO networks into $q$ angular regions of equal size are becoming "homogeneous" when the system size is increased, and the considered fluctuations across the possible different choices for $\alpha$ and across different network samples are expected to vanish in the $N\rightarrow\infty$ limit.

\section{Calculation of the expected modularity $\bar{Q}$}
In this section, we explain in detail how the expected value of the modularity can be calculated in PSO networks for a uniform partition scheme. First, in Sect.~\ref{sect:exp_mod_uniform_part_scheme} we provide general considerations on $\bar{Q}$ and formalise it as a function of the expected internal node degrees. Then, in Sect.~\ref{sect:exp_num_intra_comm_links_pso} we express the expected internal degrees with the model parameters. Sect.~\ref{sect:C1intro} deals with the $q$-dependence of $\bar{Q}$, and we introduce here the parameter $C_1$, the $N$-dependence of which is analysed in Sect.~\ref{sect:C1_N_dep}. Finally, Sect.~\ref{sect:mod_at_fix_q} presents the behaviour of $\bar{Q}(q)$ as a function of the network size $N$.

\subsection{Expected modularity for a uniform partition scheme}
\label{sect:exp_mod_uniform_part_scheme}

Let us rewrite Eq.~(3) in the main text as follows
\begin{equation}
Q=\sum\limits_{c=1}^{q}\sum\limits_{i\in c} \frac{b_i}{2E}-\sum\limits_{c=1}^{q}\left ( \sum\limits_{i \in c}\frac{k_{i}}{2E} \right )^2.
\label{eq:rewrite_mod}
\end{equation}
Provided that each node is assigned to a community, the double sum in Eq.~(\ref{eq:rewrite_mod}) can simply be replaced with a single summation running over the whole set of nodes, yielding
\begin{equation}
Q=\sum\limits_{i=1}^{N}\frac{b_i}{2E}-\sum\limits_{c=1}^{q}\left ( \sum\limits_{i \in c}\frac{k_{i}}{2E} \right )^2.
\label{eq:rewrite_mod2}
\end{equation}
Based on the fact that the angular position of the nodes is distributed uniformly and we identified the communities as 
equally-sized circular sectors of the hyperbolic disk, the sum of the node degrees has to be equal for each community in the thermodynamic $N\to \infty$ limit. 
This assumption of homogeneous mixing  
implies that
\begin{equation}
\lim_{N\to\infty}\sum\limits_{i \in c_1}k_{i} = \lim_{N\to\infty}\sum\limits_{i \in c_2}k_{i} = \lim_{N\to\infty}\frac{1}{q}\sum\limits^{N}_{i=1}k_{i}
\label{eq:mean_field_sum}
\end{equation}
for any community $c_1$ and $c_2$, based on which Eq.~(\ref{eq:rewrite_mod2}) can 
be approximated as
\begin{align}
Q&\approx \sum\limits_{i=1}^{N}\frac{b_i}{2E}-\sum\limits_{c=1}^{q}\left ( \sum\limits^N_{i=1}\frac{k_{i}}{2Eq} \right )^2=\sum\limits_{i=1}^{N}\frac{b_i}{2E}-q\left ( \sum\limits^N_{i=1}\frac{k_{i}}{2Eq} \right )^2
\\ 
&= \sum\limits_{i=1}^{N}\frac{b_i}{2E}-\frac{1}{q}.
\label{eq:rewrite_mod3}
\end{align}
By taking the expected value of the modularity $Q$ given by Eq.~(\ref{eq:rewrite_mod3}) over different PSO networks of the same model parameters, we obtain
\begin{equation}
\bar{Q}\approx\frac{\sum\limits_{i=1}^{N}\bar{b}_i}{2E}-\frac{1}{q}, 
\label{eq:rewrite_mod4}
\end{equation}
which is the same as Eq.~(5) in the main text of the article.

\subsection{The expected number of the intra-community links of the nodes in the PSO model}
\label{sect:exp_num_intra_comm_links_pso}

In order to provide a closed-form expression for the expected value of the modularity in PSO networks, one needs to compute the expected number of intra-community links $\bar{b}_i$ for each node $i$ at the end of the network generation process, i.e. at time $t=N$. 
Since $\bar{b}_i$ is a degree-based quantity, its evaluation is similar to that of $\bar{k}_i$. Thus, in this section we first revisit the derivation of $\bar{k}_i$ by following Refs.~\cite{PSO,EPSO_HyperMap} and then, we turn to discuss in detail how to calculate $\bar{b}_i$ by applying a similar set of arguments. 

\subsubsection{Derivation of the expected degree $\bar{k}_s$ of node $s$ at the end of the network growth}
\label{sect:ksderiv}

We again emphasize that here we use the variant of the PSO model called $\mathrm{Model}_{2'}$ introduced in the Supplementary Information of Ref.~\cite{PSO}. According to 
the results shown therein, during the growth of PSO networks of $T>0$, the probability that node $t$ connects to a previously appeared node $s$ can be given by
\begin{equation}
\Pi(s,t) = \frac{1}{\pi}\int\limits_0^{\pi} \frac{1}{1+\left ( \frac{X(s,t)}{2} \Delta\theta_{st} \right )^{1/T}} \,\mathrm{d}\Delta\theta_{st}\approx\frac{2T}{\sin(T\pi)}\frac{1}{X(s,t)},
\label{eq:Pst_pso}
\end{equation}
where 
\begin{equation}
    \frac{1}{X(s,t)}=e^{-\frac{\zeta}{2}\cdot(r_{st}+r_{tt}-R_t)}=s^{-\beta} t^{-(1-\beta)} \frac{\sin(T\pi)}{2T}\frac{m(1-\beta)}{1-t^{-(1-\beta)}}.
    \label{eq:xst}
\end{equation} 
By following Refs.~\cite{PSO,EPSO_HyperMap}, Eq.(\ref{eq:Pst_pso}) can 
be rephrased as  
\begin{equation}
    \Pi(s,t) = 
    m\frac{s^{-\beta}t^{-(1-\beta)}}{I_t}
    \label{eq:EPSOmeanavgdeg}
\end{equation}
with 
$I_t=\frac{1-t^{-(1-\beta)}}{1-\beta}$. 
Based on the the form of attraction probability $\Pi(s,t)$ in Eq.~(\ref{eq:EPSOmeanavgdeg}), one can determine the expected number of connections that node $s$ establishes by time $t$, yielding
\begin{align}
    \bar{k}_s(t)&=\int\limits^{s}_{1}\Pi(i,s)\mathrm{d}i+\int\limits^{t}_{s}\Pi(s,j)\mathrm{d}j\approx m+m\int\limits^{t}_{s}\frac{s^{-\beta}j^{-(1-\beta)}}{I_j}\mathrm{d}j
    \label{eq:expecteddegpso0}
    \\
    &\approx m+\frac{m}{I_t\beta}\left(\left(\frac{s}{t}\right)^{-\beta}-1\right),
    \label{eq:expecteddegpso}
\end{align}
where we have exploited that $\int^{s}_{1}\Pi(i,s)\mathrm{d}i=m$ and $I_j\approx I_t$ for sufficiently large values of $j$ and $t$, in accordance with the approximation used in Ref.~\cite{EPSO_HyperMap}. 

Note that as $T\to 0$, the connection probability in Eq.~(\ref{eq:PSO_link_prob}) converges to a reversed Heaviside step function, that is, \begin{equation}
    \lim_{T\to 0}p(x_{st}) = \left\{ \begin{array}{lll}
1\ \text{if} \ \Delta\theta_{st}\leq\frac{2}{X(s,t)}\\0 \ \text{otherwise,}
\end{array}\right. 
\label{eq:heaviside}
\end{equation}
meaning that only nodes with $\Delta\theta_{st}\leq\frac{2}{X(s,t)}$ can establish connections between one another, but those with a probability of $1$. In this case, the probability that node $t$ connects to a previously appeared node $s$ can be given by
\begin{equation}
    \Pi(s,t) = \frac{1}{\pi}\int\limits_0^{2/X(s,t)} \,\mathrm{d}\Delta\theta_{st}=\frac{2}{\pi}\frac{1}{X(s,t)},
\label{eq:Pst_pso_T0}
\end{equation}
which can also be obtained by taking the $T\to0$ limit in the right hand side of Eq.~(\ref{eq:Pst_pso}). Note that however, $\Pi(s,t)$ in Eq.(\ref{eq:EPSOmeanavgdeg}) does not depend on $T$, therefore letting $T\to 0$ does not influence the value of $\bar{k}_s(t)$ in Eq.~(\ref{eq:expecteddegpso}) either. Finally, if we are interested in the value of $\bar{k}_s(t)$ at the end of the network growth, we simply evaluate the formula appearing in Eq.(\ref{eq:expecteddegpso}) at time $t=N$, yielding
\begin{equation}
     \bar{k}_s(t=N)\equiv\bar{k}_s\approx m+\frac{m}{I_N\beta}\left(\left(\frac{s}{N}\right)^{-\beta}-1\right).
\end{equation}

\subsubsection{Derivation of the expected internal degree $\bar{b}_s$ of node $s$ at the end of the network growth}
\label{sect:bsderiv}

As described in the main text of the article, let us assume that the entire two-dimensional hyperbolic disk is divided into $q$ number of equally-sized circular sectors that correspond to communities. 
Although the angular distance $\Delta\theta_{st}$ is distributed uniformly for the whole set of nodes, 
its distribution is no longer uniform within a given circular sector. Instead, given that two nodes $s$ and $t$ fall into the same sector $\mathcal{S}_c\,(c=1,...,q)$ 
corresponding to the angular interval $\left[\alpha+\frac{2\pi(c-1)}{q},\alpha+\frac{2\pi c}{q}\right)$, the probability density for them to have angular distance $\Delta\theta_{st}$ can be calculated as follows: 
\begin{equation}
    \varrho(\Delta\theta_{st}|s,t\in \mathcal{S}_c)=\frac{q}{\pi}-\frac{1}{2}\left(\frac{q}{\pi}\right)^2\Delta\theta_{st},
    \label{eq:intra_angular_diff}
\end{equation}
where $\Delta\theta_{st}\in\left[\alpha+\frac{2\pi(c-1)}{q},\alpha+\frac{2\pi c}{q}\right)$.  
Note that by means of rotational symmetry and the statistical equivalence of the sectors the $\alpha$ and $c$ parameters above can always be set to $\alpha=0$ and $c=1$ without any loss of generality. For a detailed derivation of Eq.~(\ref{eq:intra_angular_diff}), see Sect.~\ref{sect:angDistDistr}. 

Since communities are idenfitied as 
angular sectors on the hyperbolic disk, intra-community links correspond to the connections between nodes located in the same circular sector. Based on this, the probability that node $s$ forms an intra-community link with a new-coming node $t$ given that both nodes are inside community $c$ of width $2\pi/q$ can be written as 
\begin{equation}
    \Pi_q(s,t|s,t\in \mathcal{S}_c) =\int\limits_0^{2\pi/q} \frac{\varrho(\Delta\theta_{st}|s,t\in \mathcal{S}_c)}{1+\left ( \frac{X(s,t)}{2} \Delta\theta_{st} \right )^{1/T}}   \,\mathrm{d}\Delta\theta_{st},
    \label{eq:conditional_intra}
\end{equation} 
where $\varrho(\Delta\theta_{st}|s,t\in \mathcal{S}_c)$ is given by Eq.~(\ref{eq:intra_angular_diff}). However, there are altogether $q$ number of distinct communities; therefore, the total probability that a node pair $s,t$ shares an intra-community link in any community can be written as
\begin{equation}
    \Pi_q(s,t)=\sum\limits^{q}_{c=1}\Pi_q(s,t|s,t\in \mathcal{S}_c)P(s,t\in \mathcal{S}_c),
    \label{eq:total_law_of_probs}
\end{equation} 
where $P(s,t\in \mathcal{S}_c)=1/q^2$ denotes the probability that both node $s$ and node $t$ fall into the circular sector $\mathcal{S}_c$. Due to the statistical equivalence of the communities, each term in Eq.~(\ref{eq:total_law_of_probs}) gives the same contribution, which, along with the substitution of Eq.~(\ref{eq:conditional_intra}) into Eq.~(\ref{eq:total_law_of_probs}), yields
\begin{align}
     \Pi_q(s,t) &= \frac{1}{q}\int\limits_0^{2\pi/q} \frac{\frac{q}{\pi}-\frac{1}{2}\left(\frac{q}{\pi}\right)^2\Delta\theta_{st}}{1+\left ( \frac{X(s,t)}{2} \Delta\theta_{st} \right )^{1/T}}   \,\mathrm{d}\Delta\theta_{st}
 \\
    &
    =\frac{1}{q}\int\limits_0^{2\pi/q} \frac{\frac{q}{\pi}}{1+\left ( \frac{X(s,t)}{2} \Delta\theta_{st} \right )^{1/T}}   \,\mathrm{d}\Delta\theta_{st}
    +\frac{1}{q}\int\limits_0^{2\pi/q} \frac{-\frac{1}{2}\left(\frac{q}{\pi}\right)^2\Delta\theta_{st}}{1+\left ( \frac{X(s,t)}{2} \Delta\theta_{st} \right )^{1/T}}   \,\mathrm{d}\Delta\theta_{st}
    \nonumber \\
    &:= I_1+I_2.
    \label{eq:IoneItwo}
\end{align}

Let us evaluate $I_1$ and $I_2$ separately. The first term $I_1$ turns out to have the same form as $\Pi(s,t)$ in Eq.~(\ref{eq:Pst_pso}), that is, 
\begin{align}
    I_1 &=\frac{1}{q}\int\limits_0^{2\pi/q} \frac{\frac{q}{\pi}}{1+\left ( \frac{X(s,t)}{2} \Delta\theta_{st} \right )^{1/T}}   \,\mathrm{d}\Delta\theta_{st}
    =\frac{1}{\pi}\frac{2}{X(s,t)} \int\limits_0^{\frac{X(s,t)\pi}{q}} \frac{1}{1+y^{1/T}}   \,\mathrm{d}y
    \nonumber \\ 
    &\approx \frac{1}{\pi}\frac{2}{X(s,t)} \int\limits_0^{\infty} \frac{1}{1+y^{1/T}}   \,\mathrm{d}y
    =\frac{2T}{\sin(T\pi)}\frac{1}{X(s,t)},
    \label{eq:Ioneevaluation}
\end{align}
being valid for any $T<1$ temperature values. In Eq.~(\ref{eq:Ioneevaluation}) we have also taken advantage of the fact that for sufficiently large networks at temperatures $T<1$ the main contribution to the integral $I_1$ comes from the range of small angular distances $\Delta\theta_{st}\ll 2\pi/q$
, and consequently, the upper bound of the integral can safely be extended to infinity. 
The second term $I_2$ in Eq.~(\ref{eq:IoneItwo}) is a bit more complicated to evaluate; however, similar considerations suggest that
\begin{align}
    I_2 &=-\frac{1}{q}\int\limits_0^{2\pi/q} \frac{\frac{1}{2}\left(\frac{q}{\pi}\right)^2\Delta\theta_{st}}{1+\left ( \frac{X(s,t)}{2} \Delta\theta_{st} \right )^{1/T}}   \,\mathrm{d}\Delta\theta_{st}
    = -\left(\frac{q}{\pi X(s,t)}\right)^2\int\limits_0^{\left (\frac{X(s,t)\pi}{q}\right)^2} \frac{1}{1+y^{1/(2T)}}   \,\mathrm{d}y
    \nonumber \\
    &\approx -\left(\frac{q}{\pi X(s,t)}\right)^2\int\limits_0^{\infty} \frac{1}{1+y^{1/(2T)}}   \,\mathrm{d}y=-\frac{q}{\pi} \frac{2T}{\sin (2T\pi)} \frac{1}{X^2(s,t)},
    \label{eq:Itwoevaluation}
\end{align}
where we have used the change of variables with a new variable defined as ${y=\left(\frac{X(s,t)}{2}\Delta\theta_{st}\right)^2}$. Using Eq.~(\ref{eq:heaviside}), one can show that the $T=0$ case is again well-defined. Taking the $T\to0$ limit in Eqs.~(\ref{eq:Ioneevaluation}) and (\ref{eq:Itwoevaluation}) yields
\begin{equation}
\lim_{T\to 0}I_1=\frac{2}{\pi}\frac{1}{X(s,t)}
\end{equation}
and
\begin{align}
    \lim_{T\to 0}I_2=-\frac{q}{\pi^2} \frac{1}{X^2(s,t)},
\end{align}
respectively. Nevertheless, it is important to note 
that the approximation in Eq.~(\ref{eq:Itwoevaluation}) is no longer applicable for $T\geq 1/2$ values, since in such case the corresponding integral becomes divergent.

Finally, combining Eq.~(\ref{eq:IoneItwo}) with Eqs.~(\ref{eq:Ioneevaluation}) and (\ref{eq:Itwoevaluation}) yields 
\begin{align}
    \Pi_q(s,t)=I_1+I_2\approx\frac{2T}{\sin(T\pi)}\frac{1}{X(s,t)} \left(1-\frac{q}{2\pi \cos(T\pi)}\frac{1}{X(s,t)}\right),
\end{align}
which can 
be rephrased as
\begin{align}
    \Pi_q(s,t) &=s^{-\beta} t^{-(1-\beta)}\frac{m}{I_t}\left(1-q\frac{\tan(T\pi)}{4\pi T}s^{-\beta} t^{-(1-\beta)}\frac{m}{I_t}\right)
    \\
    &=\Pi(s,t)-q\frac{\tan(T\pi)}{4\pi T}s^{-2\beta} t^{-2(1-\beta)}\frac{m^2}{I^2_t}
    \\
    &=\Pi(s,t)-q\frac{\tan(T\pi)}{4\pi T}\Pi^2(s,t),
\label{eq:piq}
\end{align}
In the above derivation, we used the formulae of $\frac{1}{X(s,t)}$ and $\Pi(s,t)$ defined by Eqs.~(\ref{eq:xst}) and~(\ref{eq:EPSOmeanavgdeg}), respectively.Analogously to Eq.~(\ref{eq:expecteddegpso0}), the expected number of intra-community links of node $s$ emerged by time $t$ can be calculated as
\begin{align}
    \bar{b}_s(t)=\int\limits^{s}_{1}\Pi_q(i,s)\mathrm{d}i+\int\limits^{t}_{s}\Pi_q(s,j)\mathrm{d}j.
    \label{eq:expectedintradegpso}
\end{align}
The first term in Eq.~(\ref{eq:expectedintradegpso}) can be simplified to
\begin{align}
  \vphantom{a} \vphantom{a} & \int_{1}^{s}\Pi_q(i,s) \mathrm{d}i  = \int_{1}^{s}\left ( \Pi(i,s)-q\frac{\tan(T\pi)}{4T\pi}\Pi^2(i,s) \right )\mathrm{d}i
\\
&=m-q\frac{\tan(T\pi)}{4T\pi}\int_{1}^{s}i^{-2\beta}s^{-2(1-\beta)}\frac{m^2}{I^2_s} \mathrm{d}i
\\
&=m-q\frac{\tan(T\pi)}{4T\pi}\frac{s^{-1}-s^{-2+2\beta}}{1-2\beta}\frac{m^2}{I^2_s},
\label{eq:K_const}
\end{align}
where we used Eq.~(\ref{eq:piq}) in the first step, and Eq.~(\ref{eq:expecteddegpso0}) together with the definition of $m=\int^{s}_{1}\Pi(i,s)\mathrm{d}i$ in the second step. 
Since the second term in Eq.~(\ref{eq:K_const}) is a decreasing function of $s$, in sufficiently large networks for the majority of the nodes the $\int^{s}_{1}\Pi_q(i,s)\mathrm{d}i$ integral is close to $m$. For the sake of simplicity, in the following, we extend this approximation for all nodes and replace the first term in Eq.~(\ref{eq:expectedintradegpso}) by $m$.

\begin{figure}[hbt]
    \centering
    \includegraphics[width=0.8\textwidth]{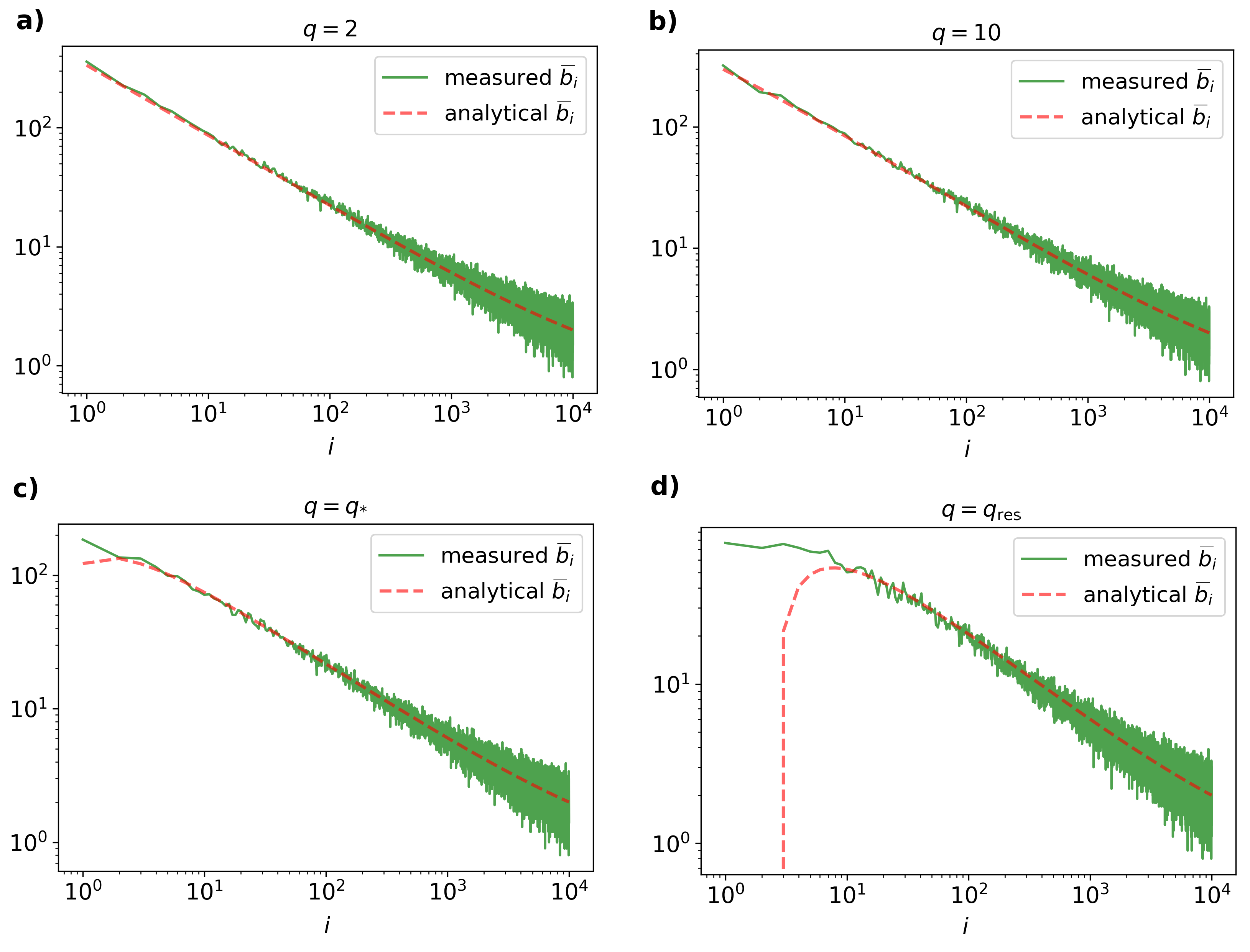}
    \caption{ {\bf The expected number of intra-community links $\bar{b}_i$ in PSO networks as a function of the node index $i$ for different values of the number of communities $q$.} The solid green curves correspond to the number of intra-community links obtained by averaging over $10$ PSO networks generated independently with 
    parameters $\zeta=1$, $N=10000$, $m=2$, $\beta=0.6$ and $T=0.1$, while the red dashed lines show the analytic prediction given by Eq.~(\ref{eq:averagebs}) for the same model parameters. The 
    number of communities used for creating panel \textbf{c} and \textbf{d} at this parameter setting were $q_{*}=48$ and $q_{\rm res}=133$, respectively. 
}
    \label{fig:bi_app_ex}
\end{figure}

Furthermore, substituting Eq.~(\ref{eq:piq}) into the second term of Eq.~(\ref{eq:expectedintradegpso}) yields
\begin{align}
    \bar{b}_s(t)&\approx m+\int\limits^{t}_{s}\Pi(s,i)\mathrm{d}i-qm^2\frac{\tan(T\pi)}{4\pi T}\int\limits^t_{s} \frac{s^{-2\beta}i^{2\beta-2}}{I^2_i}\mathrm{d}i
    \\
    &=\bar{k}_s(t)-qm^2\frac{\tan(T\pi)}{4\pi T}\int\limits^t_{s} \frac{s^{-2\beta}i^{2\beta-2}}{I^2_i}\mathrm{d}i
    \\
    &\approx \bar{k}_s(t)-qm^2\frac{\tan(T\pi)}{4\pi T}\frac{s^{-2\beta}}{I^2_t}\int\limits^t_{s}  i^{2\beta-2}\mathrm{d}i
    \\
    & =\bar{k}_s(t)-qm^2\frac{\tan(T\pi)}{4\pi T}\frac{\left(\left(\frac{s}{t}\right)^{1-2\beta}-1\right)}{s(2\beta-1)I^2_t},
\end{align}
where the same approximation has been utilised as in the case of Eq.~(\ref{eq:expecteddegpso}). In terms of modularity, we are specifically interested in the value of $\bar {b}_s(t)$ for each node $s=1,...,N$ at the end of the network generation process, i.e. at $t=N$, which simply reads as
\begin{align}
    \bar{b}_s(t=N)\equiv\bar{b}_s\approx\bar{k}_s-qm^2\frac{\tan(T\pi)}{4\pi T}\frac{\left(\left(\frac{s}{N}\right)^{1-2\beta}-1\right)}{s(2\beta-1)I^2_N}.
    \label{eq:averagebs}
\end{align}
Hereinafter, for the sake of notational simplicity, the argument of $\bar{b}_s(t)$ is always omitted when being evaluated at $t=N$. As an illustration, in Fig.~\ref{fig:bi_app_ex} we show the measured values of $\bar{b}_s$ in PSO networks as a function of the node index $s$ for different values of the number of communities $q$ in comparison with the analytical prediction given by Eq.~(\ref{eq:averagebs}). According to the results, in the $q$ regime of interest ($q\leq q_{\rm res}$) our approximation works well.

\subsubsection{Distribution of angular distances within a circular sector}
\label{sect:angDistDistr}

As it is discussed in the previous section, the angular distance of the nodes is not uniform within a circular sector, but instead follows a linearly decreasing form given by Eq.~(\ref{eq:intra_angular_diff}). In the proof of this statement, because of the statistical equivalence of the communities, it is sufficient to consider only one circular sector, namely e.g. the one that corresponds to the angular interval $\left[\alpha,\alpha+\frac{2\pi}{q}\right)$. Due to the rotational invariance that is statistically valid for the system, $\alpha$ can be set to $0$ in the proof.

First, let us examine the cumulative distribution function $F_{\Delta\theta_{st}}(\chi)$ of the angular distances inside the chosen angular sector given by the interval $\left[0,\frac{2\pi}{q}\right)$. By definition, $F_{\Delta\theta_{st}}(\chi)$ denotes the probability that the value of the angular distance $\Delta\theta_{st}$ is less than $\chi$, given that nodes $s$ and $t$ both belong to the chosen sector, i.e. $\theta_s\in\left[0,\frac{2\pi}{q}\right)$ and $\theta_t\in\left[0,\frac{2\pi}{q}\right)$. 
\begin{figure}[hbt]
    \centering
    \includegraphics[width=0.5\textwidth]{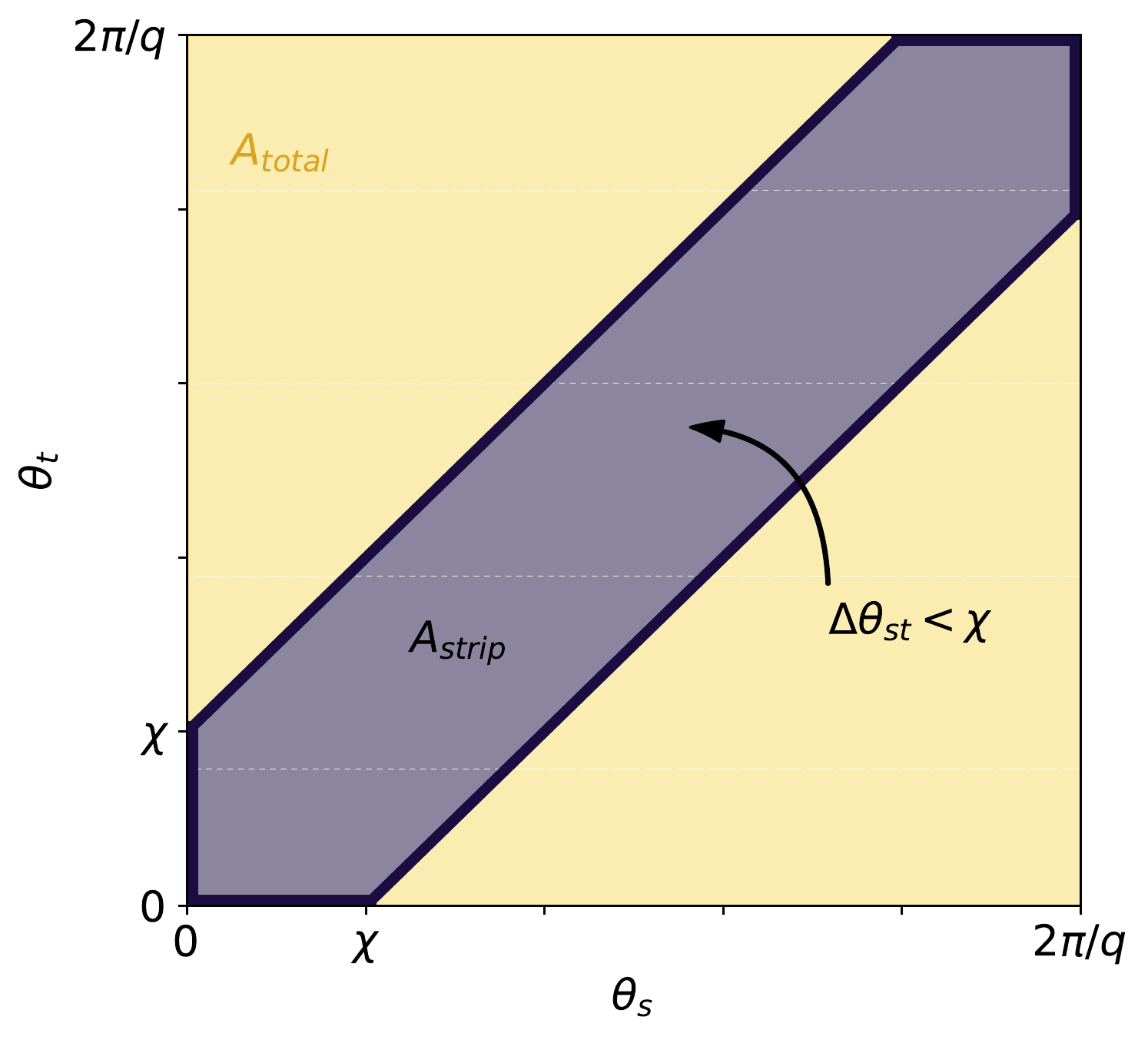}
    \caption{ {\bf 
    Graphical solution for calculating the distribution of the nodes' angular distances inside a given circular sector.} The yellow region represents all possible values of the angular coordinates $\theta_s,\theta_t$ of two arbitrary nodes $s,t$ conditioned on that they fall into the same angular sector with a central angle $2\pi/q$. The grey strip shows those combinations of $\theta_s$ and $\theta_t$ values, where $\Delta\theta_{st}<\chi$.
    }
    \label{fig:geometric_probs}
\end{figure}
Using Fig.~\ref{fig:geometric_probs}, $F_{\Delta\theta_{st}}(\chi)$ can be determined in a purely geometric way. The grey shaded strip in Fig.~\ref{fig:geometric_probs} covers the set of points where $\Delta\theta_{st}=\pi-|\pi-|\theta_s-\theta_t||=|\theta_s-\theta_t|<\chi$ holds, whereas the points of the whole square represent all possible values of the angular coordinates that nodes $s$ and $t$ can have inside the given circular sector. Since $\theta_s$ and $\theta_t$ are distributed uniformly, $F_{\Delta\theta_{st}}(\chi)$ can be calculated as the area of the grey strip $A_{\text{strip}}$ divided by the total area of the yellow square $A_{\text{total}}$ in Fig.~\ref{fig:geometric_probs}, that is,
\begin{equation}
    F_{\Delta\theta_{st}}(\chi)=\frac{A_{\text{strip}}(\chi)}{A_{\text{total}}(\chi)}=\frac{\left(\frac{2\pi}{q}\right)^2-(\frac{2\pi}{q}-\chi)^2}{\left ( \frac{2\pi}{q} \right )^2}.
    \label{eq:geometric_probs1}
\end{equation}
Now taking the derivative of Eq.~(\ref{eq:geometric_probs1}) with respect to $\chi$, we obtain the corresponding probability density function, yielding 
\begin{equation}
    \frac{\mathrm{d}}{\mathrm{d}\chi}F_{\Delta\theta_{st}}(\chi)=2\left ( \frac{q}{2\pi}\right )^2\left(\frac{2\pi}{q}-\chi\right)= \frac{q}{\pi}-\frac{1}{2}\left(\frac{q}{\pi}\right)^2\chi,
    \label{eq:geometric_probs2}
\end{equation}
from which we immediately recover Eq.~(\ref{eq:intra_angular_diff}).

\subsection{Expected modularity as a function of the number $q$ of circular sectors and the introduction of the $C_1$ parameter}
\label{sect:C1intro}

Let us now turn to the $q$-dependence of $\bar{Q}$. This can be determined by plugging Eq.~(\ref{eq:averagebs}) into Eq.~(\ref{eq:rewrite_mod4}), which yields
\begin{align}
\bar{Q}(q)&\approx \frac{\sum\limits^{N}_{i=1} \bar{k}_i-\frac{qm^2}{(2\beta-1)I^2_N}\frac{\tan(\pi T)}{4\pi T}\sum\limits^{N}_{i=1}\frac{1}{i}\left ( \left ( \frac{i}{N} \right )^{1-2\beta} -1\right )}{2E}-\frac{1}{q}
\\
&=1-q\frac{m^2}{2E(2\beta-1)I^2_N}\frac{\tan(\pi T)}{4\pi T}\sum\limits^{N}_{i=1}\frac{1}{i}\left ( \left ( \frac{i}{N} \right )^{1-2\beta} -1\right )-\frac{1}{q}
\\
&=1-C_1q-\frac{1}{q},
\label{eq:modularity_as_afunc_of_q}
\end{align}
where $C_1$ is defined as
\begin{equation}
C_1\equiv\frac{m^2}{2E(2\beta-1)I^2_N}\frac{\tan(\pi T)}{4\pi T}\sum\limits^{N}_{i=1}\frac{1}{i}\left ( \left ( \frac{i}{N} \right )^{1-2\beta} -1\right ),
\end{equation}
or, using $E\approx mN$, as
\begin{equation}
C_1=\frac{m}{2(2\beta-1)NI^2_N}\frac{\tan(\pi T)}{4\pi T}\sum\limits^{N}_{i=1}\frac{1}{i}\left ( \left ( \frac{i}{N} \right )^{1-2\beta} -1\right ).
\label{eq:C1_form}
\end{equation}

\subsection{The dependence of $C_1$ on the system size $N$}
\label{sect:C1_N_dep}

\begin{figure}[hbt]
    \centering
    \includegraphics[width=0.8\textwidth]{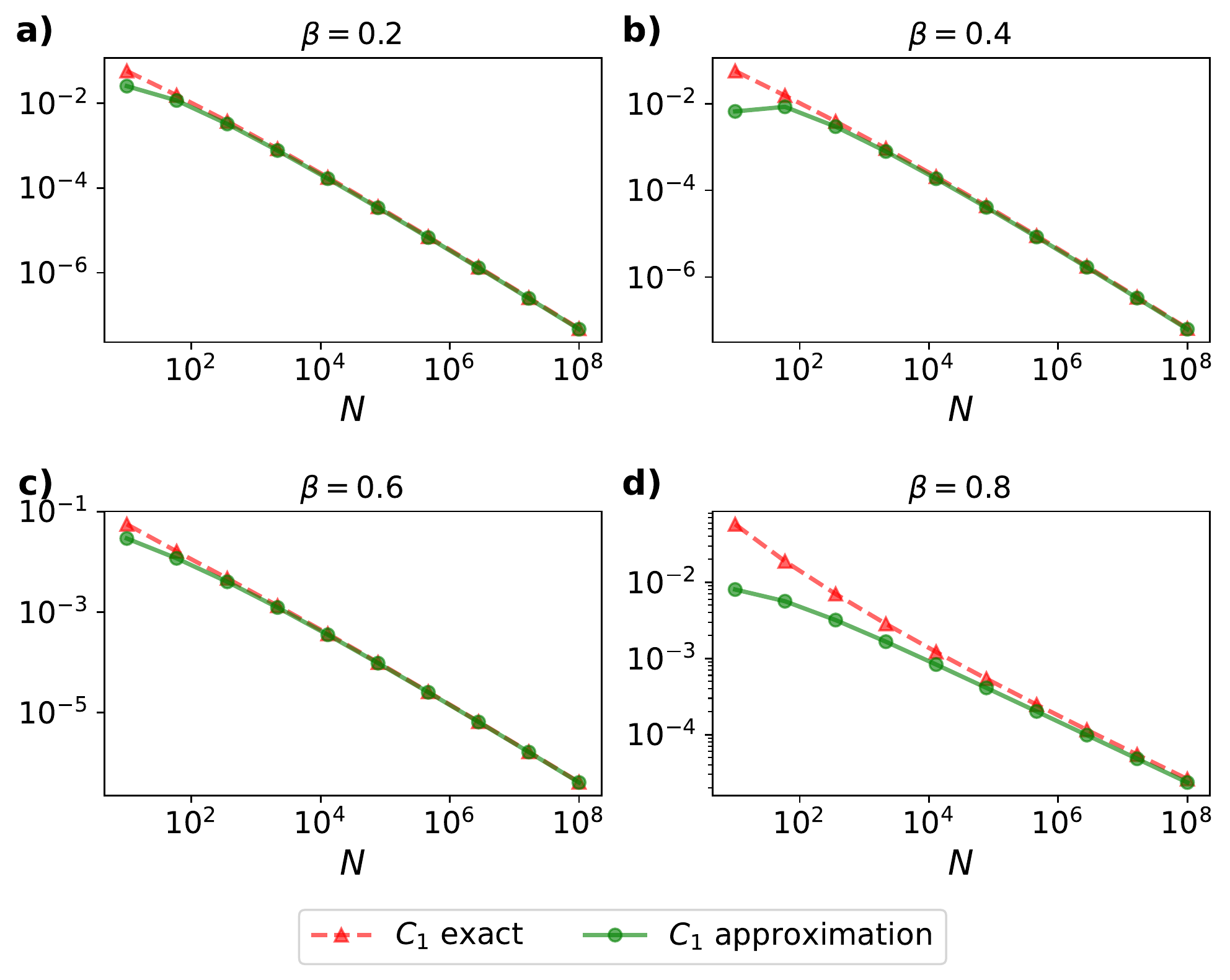}
    \caption{ {\bf Comparison of the exact and the approximate values of the parameter $C_1$ as a function of the system size $N$ for different values of the popularity fading parameter $\beta$.} In each panel (corresponding to a given value of $\beta$) we show the $C_1$ values obtained from Eq.~(\ref{eq:C1_form}) with a red dashed line, whereas the approximating form of $C_1$ defined by Eq.~(\ref{eq:approx_C1}) is displayed with a green solid line. We used $\zeta=1$, $m=2$ and $T=0.1$ in each panel.
    }
    \label{fig:C1_app_ex}
\end{figure}

Many key quantities discussed in the main text of this article -- including e.g. $q_{*}$ or $q_{\text{res}}$ -- are strongly related to the value of $C_1$. Therefore, it would be more convenient to re-express $C_1$ in a simplified form. Although it is impossible to analytically evaluate the summation in Eq.~(\ref{eq:C1_form}) for arbitrary values of $N$, fair approximations can still be done in the large network size limit, i.e. when $N\gg 1$. Using an integral approximation with a midpoint rule in Eq.~(\ref{eq:C1_form}), for $\beta\neq 1$ one obtains
\begin{align}
    \sum^N_{i=1}i^{-2\beta}&=N^{1-2\beta} \sum^N_{i=1}\frac{1}{N}\left(\frac{i}{N}\right)^{-2\beta}
    \approx N^{1-2\beta}\left(\int\limits^{1}_{1/N}x^{-2\beta}\mathrm{d}x+\frac{N^{2\beta-1}+1/N}{2}\right)
    \nonumber \\
    & = \frac{N^{1-2\beta}-1}{1-2\beta}+\frac{1+N^{-2\beta}}{2},
    \label{eq:midpoint_approx}
\end{align}
based on which $C_1$ can be simplified to
\begin{align}
    C_1&= \frac{m}{2}\frac{\tan(\pi T)}{4\pi T} \frac{\sum\limits^{N}_{i=1}\frac{1}{i}\left ( \left ( \frac{i}{N} \right )^{1-2\beta} -1\right )}{NI^2_N(2\beta-1)}
    =\frac{m}{2}\frac{\tan(\pi T)}{4\pi T} \frac{N^{2\beta-1}\sum\limits^{N}_{i=1}i^{-2\beta}-\sum\limits^{N}_{i=1}i^{-1}}{NI^2_N(2\beta-1)} \nonumber
    \\
    & \approx \frac{m}{2}\frac{\tan(\pi T)}{4\pi T} \frac{\frac{1}{1-2\beta}+\left(\frac{1}{2}-\frac{1}{1-2\beta}\right)N^{2\beta-1}-\ln N}{N\frac{2\beta-1}{(1-\beta)^2}} ,
    \label{eq:approx_C1}
\end{align}
where we used that $I_N=\frac{1-N^{-(1-\beta)}}{1-\beta}\approx \frac{1}{1-\beta}$ and $\sum^N_{i=1}i^{-1}\approx \ln N$ in the thermodynamic limit $N\to\infty$. The validity of this approximation is supported by Fig.~\ref{fig:C1_app_ex}, where we show the value of $C_1$ as a function of the number of nodes $N$ according to Eq.~(\ref{eq:C1_form}) and also its approximated form appearing in Eq.~(\ref{eq:approx_C1}).

It is worth remarking that for different values of the popularity fading parameter $\beta$, $C_1$ in Eq.~(\ref{eq:approx_C1}) behaves in a slightly different manner. Further simplifications in the thermodynamic limit suggest that
\begin{align}
C_1\approx \frac{m\tan(\pi T)}{8\pi T}\cdot\left\{
    \begin{array}{lll}
    \frac{(1-\beta)^2}{1-2\beta}\frac{\ln N}{N}, &\text{if} &\beta<\frac{1}{2},
    \\
    \frac{\ln^2N}{N}, &\text{if} &\beta=\frac{1}{2} , 
    \\
    \frac{2(\beta+1)(1-\beta)^2}{2(2\beta-1)} N^{2\beta-2}, &\text{if} &\frac{1}{2}<\beta<1,
    \end{array}\right.
    \label{eq:C1_N_dep}
\end{align}
where the $\beta=\frac{1}{2}$ case can be verified by applying the L'Hôpital's rule in Eq.~(\ref{eq:approx_C1}). 
For $\beta$ equal to $1$, $I_N$ in Eq.~(\ref{eq:C1_form}) diverges as $\ln N$, therefore it should be handled as a separate case. A quick calculation reveals that for $\beta=1$, the parameter $C_1$ decays slower than any power law, more precisely,
\begin{align}
    C_1(\beta=1)&\approx \frac{m\tan(\pi T)}{8\pi T}\frac{N\frac{\pi^2}{6}-\ln N}{N\ln^2 N}\approx \frac{m\tan(\pi T)}{8\pi T}\frac{\pi^2}{6\ln^2 N},
    \label{eq:C1_N_dep_beta1}
\end{align}
where we used the fact that $\sum^{N}_{i=1}i^{-2}\approx\frac{\pi^2}{6}$ for large values of $N$. Although $C_1$ can display fundamentally different types of scaling with $N$, it can easily be shown that for any values of the $\beta$ parameter, $C_1$ is a decreasing function of $N$ and 
\begin{equation}
    \lim_{N\to\infty} C_1 = 0,
    \label{eq:C1_goes_to_0}
\end{equation}
as a result of which further calculations in the thermodynamic limit become considerably easier.

In the Sects.~\ref{sect:optCommNum} and \ref{sect:resLim}, we derive two key quantities, namely the optimal $q_{*}$ value at which the modularity is maximal and the so-called resolution limit $q_{\text{res}}$. Moreover, we are going to discuss in detail how these quantities are related to the parameter $C_1$.

\subsection{Expected modularity at a fix value of $q$ as a function of the network size $N$}
\label{sect:mod_at_fix_q}
\begin{figure}[hbt]
\begin{center}
\includegraphics[width=0.75\textwidth]{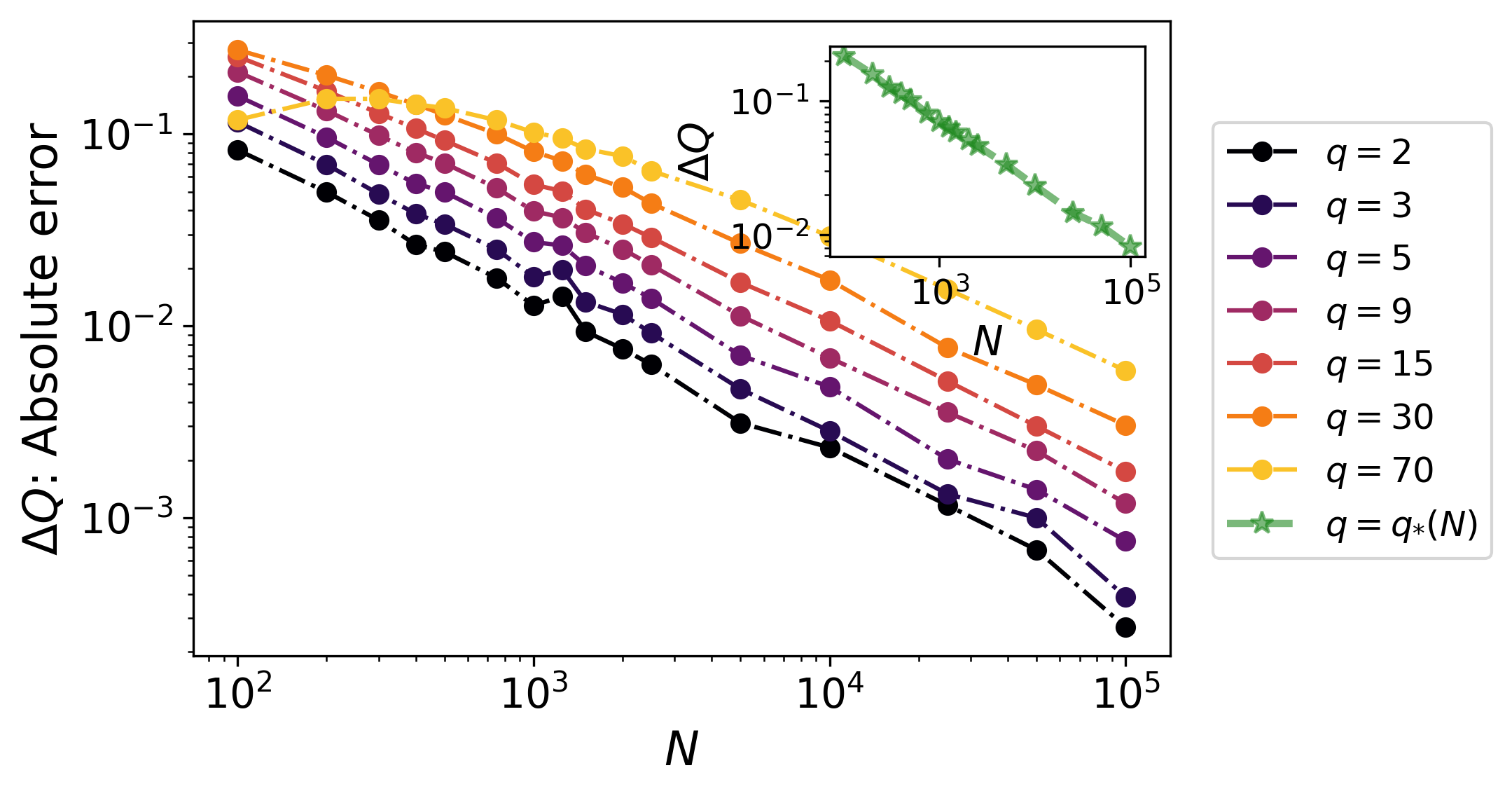}
\caption{{\bf The absolute error of the analytic modularity as a function of the network size $N$.}  The expected modularity $\bar{Q}(q)$ from Eq.~(7) of the main text is compared to the measured $Q(q)$ averaged over PSO networks of $\zeta=1$, $m=2$, $\beta=0.6$, $T=0.1$ that were divided into $q$ equally sized communities according to the angular coordinates in the native disk. The depicted absolute error is defined as ${\Delta Q=\bar{Q}(q)-\left< Q(q)\right>}$. The sample size decreases with $N$ from $10000$ at $N=10^2$ to $50$ at $N=10^5$. The inset shows $\Delta Q$ at the optimal ($N$-dependent) $q_{*}$ value, where the expected modularity is maximal.}
\label{fig:sup_abs_errors} 
\end{center}
\end{figure}

As a supplement to our analysis provided in the main text, in Fig.~\ref{fig:sup_abs_errors} we show the absolute error $\Delta Q$ with respect to the average modularity measured in PSO networks generated using the same parameter settings as in the case of Fig.~1. According to the results, the absolute error also shows a decreasing tendency with the system size $N$.

Let us provide a particularly simple limit, namely when $q$ is kept at a fix value, 
but the size of the system $N$ goes to infinity. 
According to Eqs.~(\ref{eq:modularity_as_afunc_of_q}) and (\ref{eq:C1_goes_to_0}), the expected modularity in this case is given by
\begin{equation}
    \lim_{N\to\infty}\bar{Q}(q)=1-\frac{1}{q},
\end{equation}
which is exactly the same as Eq.~(11) in the main text.

\section{Optimal number of communities}
\label{sect:optCommNum}
The optimal  $q_{*}$ is defined as the $q$ value that maximizes $\bar{Q}(q)$, i.e.
\begin{equation}
q_{*}=\max\limits_{q}\bar{Q}(q), 
\end{equation}
where $\bar{Q}(q) \approx 1-C_1q-\frac{1}{q}$ 
in accordance with Eq.~(\ref{eq:modularity_as_afunc_of_q}).
Treating now $q$ as a continuous parameter one can simply compute $q_{*}$ by setting the derivative of $\bar{Q}(q)$ to zero, which yields 
\begin{equation}
    \left. \frac{\mathrm{d}\bar{Q}(q)}{\mathrm{d}q} \right|_{q^{*}}= C_1-\frac{1}{q_{*}^{2}}=0.
    \label{eq:derivative}
\end{equation}
From Eq.~(\ref{eq:derivative}), we recover one of the main results of the paper
\begin{equation}
    q_{*}=C^{-1/2}_1.
    \label{eq:qstar_C1}
\end{equation}
Utilising Eqs.~(\ref{eq:C1_N_dep}) and (\ref{eq:C1_N_dep_beta1}) allows us to express how the optimal number of communities $q_{*}$ depends on the system size $N$, namely at $N\to\infty$ 
\begin{align}
q_{*}(N)\approx \left(\frac{8\pi T}{m\tan(\pi T)}\right)^{\frac{1}{2}}\cdot\left\{
    \begin{array}{lll}
    \frac{(1-2\beta)^{\frac{1}{2}}}{(1-\beta)}\left(\frac{N}{\ln N}\right)^{\frac{1}{2}}, &\text{if} &\beta<\frac{1}{2},
    \\
    \frac{N^{\frac{1}{2}}}{\ln N}, &\text{if} &\beta=\frac{1}{2} , 
    \\
    \frac{(2\beta-1)^{\frac{1}{2}}}{(\beta+1)^{\frac{1}{2}}(1-\beta)} N^{1-\beta}, &\text{if} &\frac{1}{2}<\beta<1,
    \\
    \frac{6^{\frac{1}{2}}}{\pi}\ln N, &\text{if} &\beta=1.
    \end{array}\right.
    \label{eq:q_opt_as_afunc_N}
\end{align}
The above results are concisely summarised in the second row of Table~I in the main text.

\subsection{Maximum value of the modularity}
\label{sect:max_val_of_mod}
The maximum value of the expected modularity is obtained by plugging Eq.~(\ref{eq:qstar_C1}) into Eq.~(\ref{eq:modularity_as_afunc_of_q}), yielding
\begin{equation}
    \bar{Q}(q_{*})=1-C_1q_{*}-\frac{1}{q_{*}}=1-\frac{2}{q_{*}}=1-2C^{1/2}_1,
    \label{eq:SI_max_modularity}
\end{equation}
which, by means of Eq.~(\ref{eq:C1_goes_to_0}), goes to $1$ in the thermodynamic limit $N\to\infty$. 

Based on Eqs.~(\ref{eq:SI_max_modularity}) and Eq.~(\ref{eq:C1_N_dep}), one can also calculate how quickly the modularity reaches $1$ as a function of $N$. 
The corresponding results are summarised in the fourth row of Table~I in the main text.

\section{Resolution limit}
\label{sect:resLim}
It is a well-known result that the maximisation of the modularity in sufficiently large networks would fail to resolve smaller communities~\cite{Fortunato_Barthelemy_resolution_limit_2007}. This can be quantified by the resolution limit providing a lower bound for the number of internal links in single modules below which merging any pair of connected communities will certainly increase the value of $Q$. In more precise terms, only those communities can be resolved by the modularity 
which have at least $l$ number of internal links with $l$ satisfying
\begin{equation}
    l>l_{\text{res}}=(E/2)^{1/2}. 
    \label{eq:resolution_limit_def}
\end{equation}
The above criterion can naturally be translated to the language of $q$, yielding 
\begin{equation}
     l\approx \frac{\sum\limits^N_{i=1}\bar{b}_{i}(q)}{2q}>\frac{\sum\limits^N_{i=1}\bar{b}_{i}(q_{\text{res}})}{2q_{\text{res}}} \equiv (E/2)^{1/2}, \label{eq:resolution_limit_equivalent}
\end{equation}
where $q_{\text{res}}$ specifies an upper limit on the number of partitions; using values of $q$ larger than $q_{\text{res}}$ would result in the communities being certainly unresolvable. By means of Eq.~(\ref{eq:averagebs}), the second part of Eq.~(\ref{eq:resolution_limit_equivalent}) can be rewritten as
\begin{equation}
    \frac{\sum\limits^N_{i=1}\bar{b}_{i}(q_{\text{res}})}{2q_{\text{res}}}\approx\frac{\sum\limits^N_{i=1}\bar{k}_{i}-2EC_1q_{\text{res}}}{2q_{\text{res}}}= \frac{2E(1-C_1q_{\text{res}})}{2q_{\text{res}}}=(E/2)^{1/2},
\label{eq:resolution_reform}
\end{equation}
which, after rearrangement, yields
\begin{align}
    q_{\text{res}} = \frac{(2E)^{1/2}}{1+C_1(2E)^{1/2}}=\frac{(2E)^{1/2}}{1+\frac{(2E)^{1/2}}{q^2_{*}}}.
    \label{eq:qres}
\end{align}
Combining this result with the inequality in Eq.~(\ref{eq:resolution_limit_equivalent}), one can obtain an equivalent reformulation of the resolution limit in Eq.~(\ref{eq:resolution_limit_def}) given by
\begin{align}
    \frac{(2E)^{1/2}}{1+C_1(2E)^{1/2}} = q_{\text{res}}>q,
\end{align}
which defines a reasonable upper limit on the number of communities $q$ to be used in our analysis.

\begin{figure}[hbt]
\begin{center}
\includegraphics[width=0.9\textwidth]{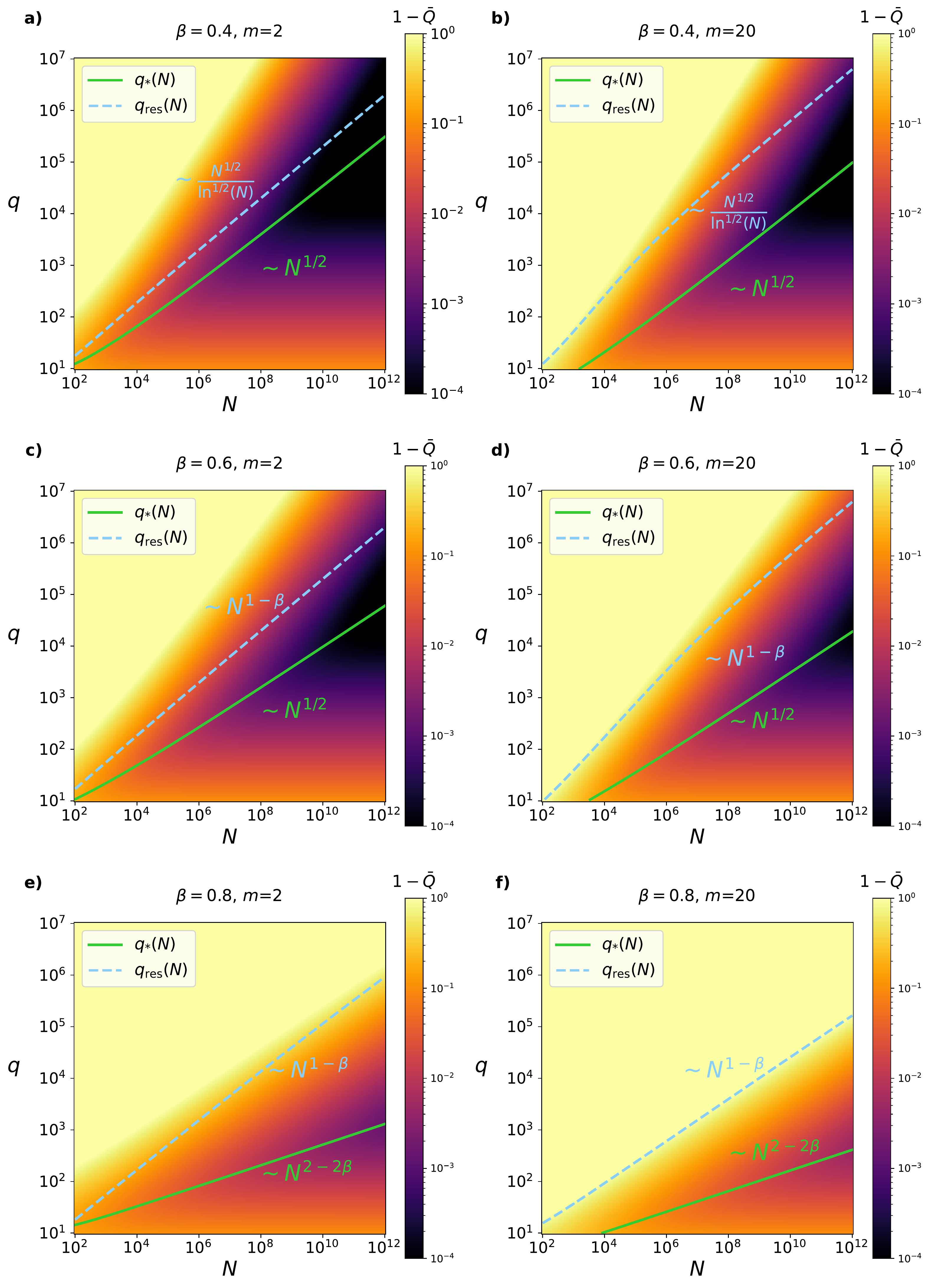}
\caption{\label{fig:multi_heatmap} {\bf Average modularity as a function of $N$ and $q$ for different values of $\beta$ and $m$.} The panels \textbf{a}–\textbf{f} display the results in a fashion similar to that of Fig.~2 in the main text: each of them shows $1-\bar{Q}$ with the help of a colormap, 
whereas continuous and dashed lines show the $q_*(N)$ and the $q_{\rm res}(N)$ curves, respectively. As indicated by the panel titles, the popularity fading parameter $\beta$ is constant across the rows (i.e. in panels \textbf{a}–\textbf{b}, \textbf{c}–\textbf{d}, \textbf{e}–\textbf{f}), while the $m$ parameter is kept at a fix value alongside the columns of the panels. The temperature $T$ was set to $0.1$ in each case. 
}
\end{center}
\label{fig:multiple_heatmaps}
\end{figure}

The dependence of $q_{\text{res}}$ on the system size $N$ is conjointly determined by $C_1(N)$ and $E\approx mN$. More precisely, taking the $N\to \infty$ limit in Eq.~(\ref{eq:qres}) and additionally using Eqs.~(\ref{eq:C1_N_dep}) and (\ref{eq:C1_N_dep_beta1}) yields 
\begin{align}
q_{\text{res}}(N)\sim \left\{
    \begin{array}{lll}
    N^{\frac{1}{2}}, &\text{if} &\beta<\frac{3}{4},
    \\
    N^{2-2\beta}, &\text{if} & \frac{3}{4} \leq \beta <1. 
    \\
    \ln^2 N , &\text{if} & \beta=1. 
    \end{array}\right.
    \label{eq:q_res_as_afunc_N}
\end{align}
As an illustration, in Fig.~\ref{fig:multi_heatmap} we show $1-\bar{Q}(q)$ as a function of both $N$ and $q$ for different values of $\beta$ and $m$, where $\bar{Q}(q)$ is obtained from Eq.~(\ref{eq:modularity_as_afunc_of_q}). Besides that, each panel of Fig.~\ref{fig:multi_heatmap} also displays the corresponding $q_*(N)$ and $q_{\rm res}(N)$ curves obtained from Eqs.~(\ref{eq:q_opt_as_afunc_N}) and (\ref{eq:qres}), respectively.

\subsection{Modularity at $q_{\rm res}$}
\label{sect:mod_at_qres}
In the followings, we discuss the rate at which $\bar{Q}(q_{\text{res}})$ goes to $1$ as a function of the system size $N$. 

The value of the modularity at $q_{\text{res}}$ can be expressed by plugging Eq.~(\ref{eq:qres}) into Eq.~(\ref{eq:modularity_as_afunc_of_q}) yielding
\begin{align}
    \bar{Q}(q_{\text{res}}) &= 1-\frac{C_1(2E)^{1/2}}{1+C_1(2E)^{1/2}}-\frac{1+C_1(2E)^{1/2}}{(2E)^{1/2}}
    \\
    & =1-\frac{2EC_1}{(2E)^{1/2}+2EC_1}-\frac{(1+C_1(2E)^{1/2})^2}{(2E)^{1/2}+2EC_1}.
    \label{eq:mod_at_qres}
\end{align}
Since $E\sim N$ and $C_1$ is a decreasing function of $N$, the dominant part of $\bar{Q}(q_{\text{res}})$ is always given by the first term in Eq.~(\ref{eq:mod_at_qres}). This simplification suggests that 
\begin{equation}
    1-\bar{Q}(q_{\text{res}})\approx  \frac{C_1(2E)^{1/2}}{1+C_1(2E)^{1/2}},
    \label{eq:mod_at_qres_1}
\end{equation}
which, along with the result obtained for $C_1$ in Eq.~(\ref{eq:C1_N_dep}), leads to
\begin{equation}
    1-\bar{Q}(q_{\text{res}})\approx C_1(2E)^{1/2} \sim \left\{
    \begin{array}{lll}
    \frac{\ln N}{N^{1/2}}, &\text{if} &\beta<\frac{1}{2},
    \\
    \frac{\ln^2N}{N^{1/2}}, &\text{if} &\beta=\frac{1}{2} , 
    \\
    N^{2\beta-3/2}, &\text{if} &\frac{1}{2}<\beta<\frac{3}{4},
    \end{array}\right.
    \label{eq:mod_at_qres_2}
\end{equation}
where we exploited the fact that $C_1(2E)^{1/2}$ goes to zero as $N\to \infty$ for $\beta < 3/4$, and thus, the Taylor-expansion of $\frac{x}{1+x}\approx x$ with $x=C_1(2E)^{1/2}$ can be used in Eq.~(\ref{eq:mod_at_qres_1}). Note, however, that for ${\beta \geq 3/4}$ values the above approximation does not hold since $C_1(2E)^{1/2}$ in Eq.~(\ref{eq:mod_at_qres_1}) is either divergent or converges to a non-zero value in the thermodynamic limit. More precisely, based on Eqs.~(\ref{eq:C1_N_dep}) and~(\ref{eq:mod_at_qres}) one can show that for $\beta = 3/4$ 
\begin{align}
1-\bar{Q}(q_{\text{res}})\approx \frac{K_1}{1+K_1}+ \frac{1+K_1}{(2mN)^{1/2}}\approx \frac{K_1}{1+K_1},
\label{eq:mod_at_qres_3pre}
\end{align}
where $K_1:=\frac{(2m)^{1/2}m \tan(\pi T)}{8\pi T}\frac{\left(\frac{3}{4}+1\right)\left(1-\frac{3}{4}\right)^2}{\left(\frac{3}{2}-1\right)}$ is a constant which does not depend on the system size $N$. For $\beta > 3/4$, similar considerations suggest that 
we can approximate $1-\bar{Q}(q_{\text{res}})$ as
\begin{align}
1-\bar{Q}(q_{\text{res}})\approx\frac{K_2N^{2\beta-3/2}}{1+K_2N^{2\beta-3/2}}\approx\frac{1}{1+K^{-1}_2N^{3/2-2\beta}}\approx 1 - K^{-1}_2N^{3/2-2\beta},
\label{eq:mod_at_qres_3}
\end{align}
where $K_2$ is defined as $K_2:=\frac{(2m)^{1/2}m\tan(\pi T)}{8\pi T}\frac{(\beta+1)(1-\beta)^2}{(2\beta-1)}$. In the last step above we have exploited that since $K_2N^{2\beta-3/2}\to \infty$ as $N\to\infty$, therefore the Taylor expansion $\frac{1}{1+x}\approx1-x$ with $x=K^{-1}_2N^{3/2-2\beta}$ can be utilised in Eq.~(\ref{eq:mod_at_qres_3}). The results presented herein are concisely summarised in Table~I of the main text.



\end{document}